\documentclass[manuscript, screen]{jair}

\RequirePackage[
  datamodel=acmdatamodel,
  style=acmauthoryear,
  backend=biber,
  giveninits=true,
  uniquename=init
  ]{biblatex}

\usepackage{graphicx}
\usepackage{multirow}
\usepackage{amsmath}
\usepackage{amsfonts}
\usepackage{amsthm}
\usepackage{mathrsfs}
\usepackage{booktabs} 
\usepackage{url}
\usepackage{array} 

\renewcommand\received[1]{}
\setcopyright{none}
\renewcommand\footnotetextcopyrightpermission[1]{}

\providecommand{\JAIRTrack}{} 
\renewcommand{\JAIRTrack}{}
\providecommand{\JAIRAE}{} 
\renewcommand{\JAIRAE}{}
\begin{document}

\title[The Personalization Paradox]{The Personalization Paradox: Semantic Loss vs. Reasoning Gains in Agentic AI Q\&A}

\author{Satyajit Movidi}
\authornote{Corresponding Author.}
\email{sm402@students.uwf.edu}
\affiliation{%
  \institution{University of West Florida}
  \city{Pensacola}
  \state{FL}
  \country{USA}
}

\author{Stephen Russell}
\email{russell@uwf.edu}
\affiliation{%
  \institution{University of West Florida}
  \city{Pensacola}
  \state{FL}
  \country{USA}
}

\renewcommand{\shortauthors}{Movidi \& Russell}

\begin{abstract}
\vspace{1em}
\emph{\textbf{\large Abstract}} \\[0.5em]
{\bf Background:} 
    AIVisor is an agentic retrieval augmented large language model (RAG LLM) system for student advising, designed to evaluate the complex trade-offs of personalization in question answering. 
    
    {\bf Objectives:}
    To investigate the nature and direction of performance shift when personalization is introduced to an LLM/RAG system, specifically testing the hypothesis that personalization creates critical metric-dependent trade-offs rather than uniform gains.
    
    {\bf Methods:}
    Twelve authentic advising questions, intentionally skewed toward lexical precision to create a conservative test, were used to compare ten factorial personalized and non-personalized configurations. We employed a Linear Mixed-Effects Model (LMM) to analyze performance across lexical (BLEU, ROUGE-L), semantic (METEOR, BERTScore), and RAGAS grounding measures.
    
    {\bf Results:}
    The LMM revealed a significant underlying trade-off: Personalization factors improved reasoning quality but incurred a significant negative interaction on semantic similarity (BERTScore). This semantic penalty is identified as a methodological artifact, where standard metrics penalize the very deviations from a generic reference that personalization is designed to create, exposing a fundamental flaw in current LLM evaluation systems. The fully integrated configuration achieved the highest composite gain, driven by reasoning and grounding improvements.
    
    {\bf Conclusions:}
    Personalization creates critical metric-dependent trade-offs, demonstrating that current semantic evaluation paradigms are insufficient for assessing user-specific AI responses. The findings establish a baseline and methodological framework for robust, transparent personalization in agentic AI.
\end{abstract}




\maketitle

\thispagestyle{empty} 
\pagestyle{plain}     

\section{Introduction}
The adoption of large language models and agentic artificial intelligence (AI) systems is accelerating across higher education, especially for student-facing services such as advising. Students increasingly expect on-demand guidance, yet traditional advising remains uneven in availability and places pressure on faculty and staff to meet students needs and expectations. AI-enabled advising tools promise to expand access and reduce advisor burden, a critical need given increasing advising workloads and institutional pressures \cite{Thues2024}. Many systems today rely on generic retrieval augmented generation (RAG) pipelines that provide factually correct but context-agnostic responses. Such systems rarely integrate personalization in the form of a student's academic history, goals, or constraints, creating risks of disengagement or misalignment with individual needs \cite{Lang2025}. Personalization can provide a mechanism that conditions responses based on users' intent, role, or context so system-provided answers are both right and right-for-the-user. 

Advances in AI have fostered a simultaneous promise of personalization, e.g., personal assistants, personalized healthcare, personalized marketing, and personalized education. While research on personalization in AI has expanded substantially in fields such as recommendation systems and affective computing \cite{Hasan2025, Li2023}, higher education advising has remained mostly outside of that scope. Most AI advising tools in the literature emphasize response efficiency and coverage rather than semantic alignment to a user's context or specifics. When personalization is considered in the AI or human computer interaction (HCI) literature, the emphasis tends to be on optimizing retrieval or ranking metrics instead of examining the broader human interaction required in advising. RAG techniques have improved factual accuracy and knowledge grounding \cite{Gao2023}, but the challenge of aligning answers with student intent, situated background, or personal context remains largely unexamined. Prior work on personalization in human-AI interaction emphasize the role personalization can play in producing more contextually and semantically appropriate responses \cite{Kapoor2023, Bach2023}, though this effect has not been evaluated in advising systems.

Intuitively, many believe personalization is always beneficial. It is also becoming increasingly implicated in modern AI systems. Yet, prior studies on AI advising or Q\&A have largely emphasized descriptive outcomes, leaving open the theoretical question of when personalization is appropriate and providing limited methodological treatment of how robust the personalization effects are. To investigate the implications of personalization on student advising, we developed AiVisor, a prototype agentic AI advising system incorporating RAG, which integrates a personalization capability. We hypothesize that personalization matters because it can shift the alignment of answers, and we use AiVisor to test the nature and direction of this shift across different metric families. Our response evaluation reports measured effects across similarity and semantic metrics (BLEU, ROUGE-L, METEOR, BERTScore) and also applies rigorous statistical analysis to assess the significance of observed differences. Importantly, we evaluate AiVisor on a deliberately stringent, conservative test where the question set ($n=12$) was intentionally skewed toward lexical and generic phrasing, emphasizing exact wording over open-ended synthesis. This setting constitutes a "worst-case" evaluation, designed to reveal the complex trade-offs and methodological challenges that arise when personalization deviates from a single, generic ground truth.

AiVisor distinguishes itself from prior advising and Q\&A research by making personalization the central experimental variable rather than an assumed benefit. Our study shows that the effects of personalization vary by question type, with gains in some cases and negligible or even negative impacts in others. Unlike earlier work that relied on small pilots or perception surveys, AiVisor employs multi-metric, semantic, and statistical evaluation, bridging technical system analysis with user-centered advising outcomes. The remainder of the paper proceeds as follows. We review related work on personalization and AI student advising Q\&A systems, describe the AiVisor system design and experimental methodology, and present the results. Lastly, we conclude with implications for the development of future advising systems.

\section{Background and Related Work}
Personalization has become a cornerstone of modern AI, enabling systems to tailor interactions to individual users' preferences, context, and goals \cite{Li2025}. Large language models (LLMs) in their vanilla form excel at general knowledge tasks but struggle with user-specific adaptation. For example, they have difficulty understanding an individual's emotions, style, or preferences without additional context. This gap has spurred research into personalized LLMs and RAG systems that leverage user data (profiles, history, etc.) to produce responses more relevant to each user \cite{liu2025survey}. Across domains like marketing and human-computer interaction, personalization is recognized as key to user engagement. In marketing, for instance, it is considered a key strategy for delivering relevant customer experiences, though it requires deep understanding of user needs and behaviors \cite{Chandra2022}. Similarly, AI-driven services such as chatbots and recommender systems increasingly incorporate personalization to improve user satisfaction and effectiveness. A survey by \citet{Li2025}, for example, systematically examines how personalization can be integrated at all stages of RAG, yet it also emphasizes that prior work has not comprehensively addressed personalized RAG. Most existing surveys have focused on general RAG or agents without systematically exploring their implications for personalization. In short, while personalization is widely pursued in AI, its integration with advanced frameworks like RAG is still an evolving research area.

In adjacent domains, such as chatbots and recommender systems, personalization has been studied for years. For example, \citet{Blomker2025} examine how few-shot methods can craft distinct conversational personas in service chatbots, enhancing user engagement and perceived alignment with the user. The notion of few-shot personalization of LLMs is also reflected in recent work: \citet{Kim2024} propose a system that iteratively adjusts prompts per user using ``mis-aligned responses'' to refine personalized performance. Similarly, personalization is not just a superficial feature (e.g. calling the user by their name), but involves modeling preferences, past interactions, and adapting system behavior over time. This personalization research focus can be seen in the dialog system proposed by \citet{Siddique2023}, where they illustrate a zero-shot reward-based personalization framework for task-oriented dialogue systems, enabling adaptation without per-user labeled data. Recent literature reviews also document trends in personalized AI marketing, identifying major themes like recommendation engines, customer segmentation, and real-time individualization \cite{Chandra2022,CasacaMiguel2024,Tarifi2024}.

\begin{table}[htbp]
\caption{Brief summary of adjacent domain literature}
\label{tab:ref_summary_marketing}
\centering
\renewcommand{\arraystretch}{1.2}
\begin{tabular}{>{\raggedright\arraybackslash}p{2.0cm} >{\raggedright\arraybackslash}p{3cm} p{8.5cm}}

\toprule
\textbf{Reference} & \textbf{Domain / Context} & \textbf{Methods / Approach / Limitations} \\
\midrule
\citet{Li2025} & LLM / RAG personalization & Survey of personalization techniques (pre-retrieval, retrieval, generation). Identifies gaps in systematic personalization, evaluation, and architecture integration. \\
\citet{Akiba2023} & Academic advising / ChatGPT & Empirical study of ChatGPT on advising queries. Strengths in general advice, but limitations in nuanced student-level guidance. \\
\citet{Antico2024} & RAG-based advising chatbot & Built RAG assistant, pilot test. Observed incomplete answers, hallucinations, broken citations, and user trust issues. \\
\citet{Blomker2025} & Service / customer chatbots & Few-shot persona adaptation for chatbots. Improves conversational fit. Limit: scalability, stability, and consistency of persona. \\
\citet{Kim2024} & LLM personalization & Proposed Fermi: few-shot prompt tuning using mis-aligned responses. Gains in personalization, but vulnerability to noise and misalignment. \\
\citet{Siddique2023} & Dialog systems personalization & Zero-shot reward-based personalization for dialogue. Reduces annotation needs; challenges in domain transfer and fairness. \\
\citet{Chandra2022} & Marketing / digital personalization & Review of trends: recommender systems, segmentation, dynamic content adaptation. Notes complexity beyond name insertion. \\
\citet{CasacaMiguel2024} & Consumer research & Systematic review of personalization's impact on consumer satisfaction. Explores moderating effects like privacy concerns. \\
\citet{Tarifi2024} & Marketing strategy / applied & Empirical/strategic work on personalization tactics (recommendation, dynamic content). Discusses effectiveness and limitations. \\
\bottomrule
\end{tabular}
\end{table}

\subsection{Personalization and Question \& Answer Systems}

Contemporary personalization studies often evaluate on prompts that obviously benefit from context or they measure with a single metric, risking overfitting to that measure. Few papers normalize across metrics, report paired statistics, or ask whether gains persist when questions are generic and strictly lexical. AiVisor addresses these gaps by using a worst-case, lexical-heavy suite, multi-metric z-score normalization, and paired tests with effect sizes. This combination is designed to separate true semantic improvements from superficial paraphrase and clarifies where lexical penalties appear and how large they are.

Most Q\&A benchmarks and leaderboards optimize for generic performance. Datasets such as SQuAD, MS MARCO, and Natural Questions largely omit user-specific context. While recent LLM-based systems show promise \cite{Ji2023HallucinationSurvey}, without grounding, they risk hallucination and a lack of specificity. RAG approaches \cite{Lewis2020RAG} combine retrieval with LLMs to improve factuality, but the literature still offers limited, rigorous evidence for user-centric personalization in Q\&A systems, especially under lexical stringency. This gap is often measured using standard evaluation metrics like BLEU \cite{Papineni2002BLEU}, ROUGE-L \cite{Lin2004ROUGE}, METEOR \cite{Banerjee2005METEOR}, and BERTScore \cite{Zhang2020BERTScore}, which provide complementary views of lexical and semantic overlap.

\subsection{Student Advising Systems}

\citet{Thottoli2024} conducted a literature review and found only 67 relevant papers, of several hundred search results, covering 1984-2023 on chatbots or AI in academic advising. For example, \citet{LinYu2023} performed a bibliometric analysis of educational chatbots, and \citet{Bilquise2022} proposed a bilingual advising chatbot. Their review highlighted that this research was spread across several domains (e.g., education technology, computer science, management, and information systems) and disparate topics (e.g., student engagement, FAQ answering, enrollment support, or counseling), lacking a cohesive research agenda or connections to build cumulative knowledge. They also found a lack of common or unified frameworks, and shallow empirical validation. They concluded that minimal study has been done on applying chatbots and AI for academic advising in higher education institutions, identifying this as a major research gap.

\citet{Dawood2024} provides a narrative literature review on the use of chatbots for personalized academic advising in higher education, synthesizing findings from 20 recent studies (2020-2024). This work defines personalization as tailored responses beyond generic frequently asked questions, recommender-based suggestions, and dynamic adaptation in multi-turn dialog. \citet{Dawood2024}'s review concluded that while ``personalized'' chatbots hold transformative potential for scaling advising services, especially in reducing advisor workloads, current systems remain limited in handling complex advising. However, \citet{Dawood2024} emphasizes that most existing systems claim personalization, but in practice, many fall short, delivering only generic or rule-based answers. The review highlights this as one of the field's biggest gaps.

Despite the sparsity of work found by \citet{Thottoli2024}, and the limitations found by \citet{Dawood2024}, there is some literature directly related to RAG-based LLM student advising systems. For example, \citet{Akiba2023} examined ChatGPT's potential in academic advising and remarked that this context does not appear to have been very well studied so far. They point out that much of the early discourse around tools like ChatGPT in higher education has focused on cheating or broad pedagogical impact, while student support services (like advising) were comparatively neglected. \citet{Antico2024} constructed a RAG-driven chatbot system by ingesting institutional documents and customizing GPT prompts for the University of Milano-Bicocca. They ran a small usability pilot (n=6) and found positive responses to the interface and conversational tone. However, they also observed that the system sometimes failed to provide fully accurate answers, omitted relevant document content, and generated un-clickable or broken link issues.  Another work by \citet{abdelhamid2024advisely} introduced Advisely, a GPT-4-powered academic advising platform designed to address the common challenges of traditional advising such as inconsistent guidance, limited accessibility, and high advisor workloads.  \citet{abdelhamid2024advisely} demonstrated that institution-specific knowledge integration combined with LLM capabilities can improve advising accessibility, accuracy, and consistency.

\citet{Soomro2025} examined how AI-driven academic advising systems can support student success by improving retention, enhancing academic confidence, and aligning educational choices with career pathways. Their study surveyed 1,200 university students across three institutions that had deployed AI-powered advising platforms. The AI advising systems studied by \citet{Soomro2025} combined NLP, machine learning-based recommendation engines, predictive analytics, and conversational chatbots integrated with institutional data, enabling tailored course advice and career-aligned guidance. Unlike architecture-specific implementations (e.g., agentic RAG or few-shot personalization), their focus was on the functional composition of advising platforms and students' perceived personalization benefits. This survey-based evaluation of *perceived* outcomes contrasts with the present study, which operationalizes personalization via *algorithmic* adaptation (agentic RAG) and evaluates its effects using objective performance metrics.

\citet{Luong2025} present FIT-Advisor, a chatbot-based academic advising system developed for Information Technology students. Their approach integrates natural language processing, machine learning, and recommender algorithms to assist students with training regulations, curriculum information, basic IT knowledge, and recommended course suggestions. This study demonstrated how localized AI-driven advising systems can improve accessibility and accuracy in higher education. \citet{Luong2025} highlight that FIT-Advisor should complement rather than replace human advisors, reserving complex advising cases for human expertise while offloading routine advising tasks to the chatbot.
 
Another work by \citet{tamascelli2025academic} is quite comprehensive in its contributions. This work extends an existing system based on Gemini 1 Pro and RAG over a university student handbook with an agent-based system built on GPT-4 with the text-embedding-ada-002 model. They integrated three tools: RAG (student handbook), Google Search (for out-of-scope queries), and Gmail (for human-in-the-loop escalation). Evaluation used 18  questions derived from the student handbook, scored with the RAGAs framework (answer correctness, context precision, faithfulness, entity recall, and answer relevancy). Their results showed that the agentic version outperformed the retrieval-only version across all metrics except entity recall, with strong performance in conversational flow, contextual relevance, and reliability.  Moreover, this work incorporated human-in-the-loop and hallucination mitigation capabilities. However, this effort did not address personalization or lexical and semantic concerns.
 
These individual efforts show that the idea of AI advisors is emerging, but the integration of personalization into such systems remains largely unexplored. The literature review highlights that even widely cited papers in this area cover diverse topics with no clear demarcations. In short, personalized AI advising is an under-researched area, with only nascent studies pointing to its promise. This is an important observation for academics and practitioners: unlike domains such as marketing or e-commerce where personalization techniques are relatively mature, in student advising the field is only beginning to ask fundamental questions and run early experiments.

While prior research has explored AI based advising systems, most efforts have been either architectural demonstrations such as retrieval augmented generation versus agentic designs, small scale institutional pilots such as Advisely and FIT Advisor, or survey based evaluations of perceived personalization. These studies highlight the potential of AI to scale advising, yet remain fragmented, often limited by small evaluation sets, self reported outcomes, or narrow technical scope. AiVisor differentiates itself from prior work in four key ways:
\begin{enumerate}
    \item \textbf{Personalization efficacy as the central research variable.}  
    Prior work has generally treated personalization as an assumed benefit or measured it indirectly through student surveys. AiVisor instead isolates personalization as the experimental factor, directly testing whether it improves or weakens advising responses.  
    \item \textbf{Question type sensitivity.}  
    AiVisor demonstrates that personalization does not uniformly improve outcomes. Some advising question types show strong gains, while others see little to no benefit or even negative impacts in others. This nuanced result challenges the assumption that personalization is universally positive and brings new theoretical insight to advising research.  
    \item \textbf{Methodological rigor.}  
    Whereas previous studies often relied on small scale descriptive metrics or subjective feedback, AiVisor applies a suite of lexical and semantic similarity measures including BLEU, ROUGE-L, BERTScore, and METEOR, combined with statistical testing. This provides a more robust and quantitative evaluation framework than has been used in advising research to date.  
    \item \textbf{Bridging technical and user-centered evaluation.}  
    Prior studies tend to focus either on technical architecture comparisons or on user perceptions of usefulness. AiVisor positions itself in between, offering system-level evidence of personalization effects that can inform both design choices and user-facing outcomes.  
\end{enumerate}

Together, these contributions establish AiVisor as the first advising study to rigorously and systematically evaluate the real impact of personalization, providing both methodological innovation and theoretical differentiation from prior literature. Before providing details on the experimental methodology, an overview of the system design is presented in the next section.

\section{System Design}
\begin{figure}[ht]
    \centering
    \includegraphics[width=0.85\textwidth]{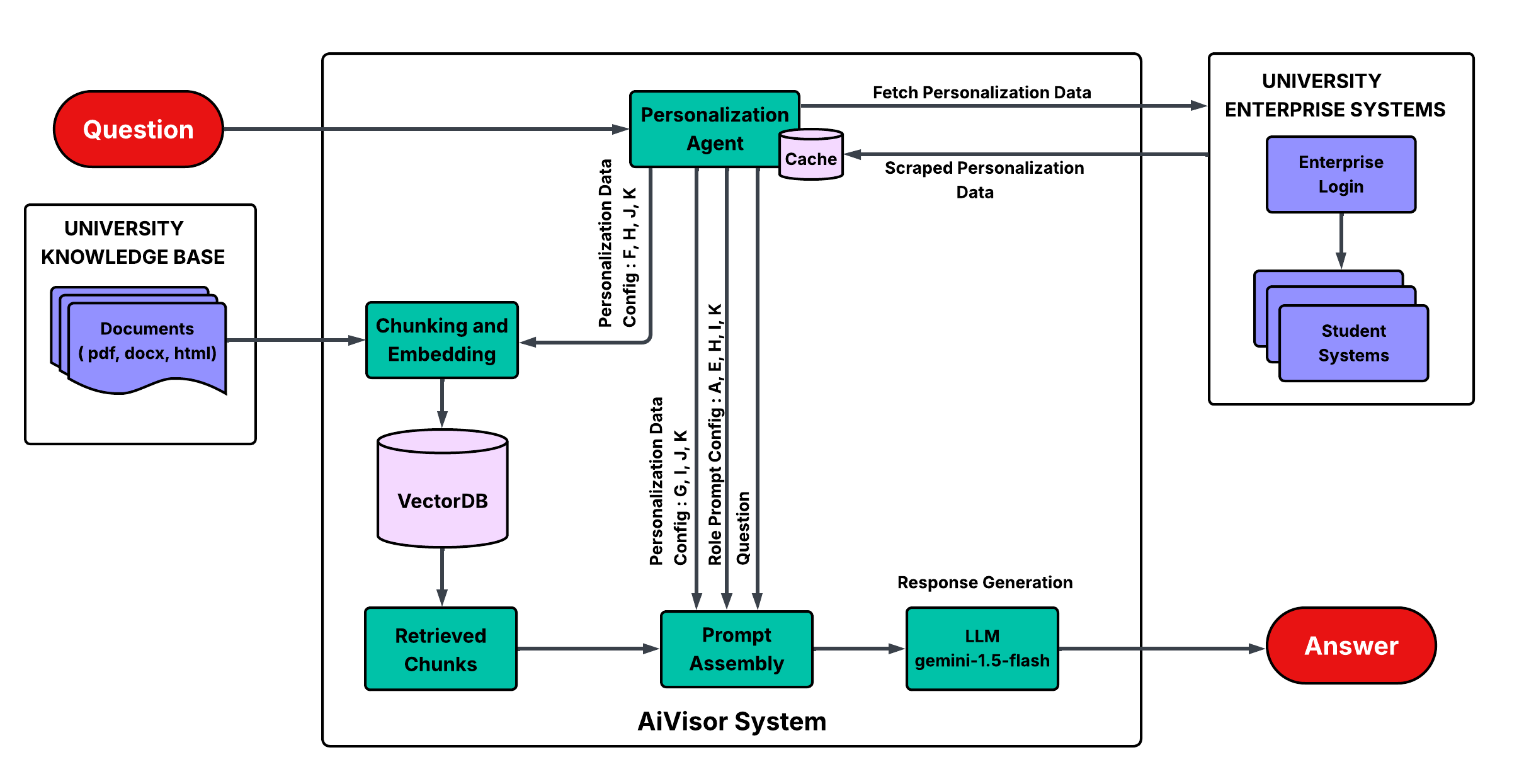}
    \caption{Block illustration of the AiVisor system with Personalization Agent, VectorDB, and Prompt Assembly.}
    \label{fig:sysdiagram}
\end{figure}

 The AiVisor system is designed around Gemini-Flash 1.5 LLM, an agent that retrieves personalization information, and a Facebook AI Similarity Search (FAISS) vector database (vectorDB) retrieval-augmented generation (RAG) pipeline. Figure~\ref{fig:sysdiagram} presents a block illustration of AiVisor.  Institutional documents (policies, catalogs, and Frequently Asked Questions - FAQs) form the core retrieval corpus, while the user-personalization conditioning agent optionally injects personalization content into either vectorDB or prompt assembly. Prompt assembly composes the LLM input, which includes a role prompt, the RAG context, personalization information, and the user's question. Some elements of the assembled prompt are included or excluded depending on the system configuration.

\subsection{Chunking Configuration} 
Documents were segmented into 10,000-character chunks with a 1,000-character overlap using a recursive text splitter. Gemini Flash 1.5 supports context windows of up to 1,048,576 tokens \cite{GoogleGeminiLongContext}, enabling the use of larger retrieval units that preserve cross-section dependencies such as definitions, tables, and cross-references. Maintaining such coherence reduces hallucinations arising from fragmenting logically connected text \cite{Zhang2025HallucinationSurvey}. Keeping quoted or cited material and its supporting context within a single retrieval unit also lowers the risk of accurate quotations being divorced from their qualifying details. In contrast, small chunks typically require large overlaps to prevent context breaks, inflating storage with near-duplicate vectors. Larger chunks, therefore, minimize both redundancy and fragmentation.

For retrieval, we used the FAISS vectorDB with the default k=4, including all retrieved chunks in the prompt assembly. While coarse chunking improves the internal coherence of policy and FAQ passages, it can also blur topical boundaries, introduce irrelevant context, and reduce retriever precision. Consequently, the larger chunk size increases background noise and may partially obscure personalization signals. Because retrieval optimization was not the central focus of this study, the chosen coarse-grained chunking (10,000-character chunks) should be viewed as a conservative methodological condition that likely attenuates any observed benefit of personalization by fragmenting contextual relevance signals.

\subsection{Experimental System Configurations} 

Ten system configurations (A-K) that varied in their use of RAG, role conditioning, and personalization were tested. Configuration B was omitted from testing to avoid any privacy concerns. Table~\ref{tab:sysconfigs} provides the details of each configuration. Configuration A served as a non-RAG web baseline, using only web search results and a role prompt for grounding, while C removed both role and personalization cues to represent a pure web-only condition. Configuration D introduced RAG with FAQs and policies, but without prompts or personalization. Hereafter, all configurations included the FAQs and policies within the RAG vector database. Building on this, E incorporated a role prompt to make retrieval role-aware, and F embedded personalization directly within the retriever. Configuration G shifted personalization to the generation stage, applying a personalization prompt after retrieval, whereas H combined role context and embedded personalization for more contextualized outputs. Configuration I employed dual-layer prompting, adding both role and personalization prompts alongside retrieved data. Configuration J created redundant personalization, applying user context in both retrieval and generation, and K represented the fully integrated system, combining role input, retriever-level personalization, and redundant prompt-level personalization. Each configuration held decoding and 0-level temperature settings constant to isolate personalization effects. These system configurations were designed to be compared in specific pairs. This pairing isolates the independent and interactive effects of role-framing, retrieval grounding, and personalization, as detailed in Table~\ref{tab:effect_pairs}.

\begin{table}[htbp]
\caption{System Configuration Summary}
\label{tab:sysconfigs}
\centering
\renewcommand{\arraystretch}{1.2}
\begin{tabular}{>{\raggedright\arraybackslash}p{0.7cm} >{\raggedright\arraybackslash}p{4.3cm} p{8.0cm}}
\toprule
\textbf{Conf\.} & \textbf{Configuration} & \textbf{Description} \\ 
\midrule
A & Web Baseline (No RAG) & Uses web results with a role prompt; serves as the non-RAG reference system. \\ 
B & Web + Personalization & Adds a personalization prompt to the web baseline for user-context adaptation. \\ 
C & Web-Only (No Role/Personalization) & Generates responses directly from the user query without role or personalization context. \\ 
D & VectorDB-Only RAG Baseline & Retrieves from internal sources (FAQ, policy, KB) with no prompts or personalization. \\ 
E & RAG with Role Prompt & Adds a role prompt to the RAG baseline for contextual, role-aware retrieval. \\ 
F & Embedded Personalization in Retriever & Integrates personalization within VectorDB retrieval, tailoring results at the retriever level. \\ 
G & RAG with External Personalization Prompt & Applies personalization at the generation stage, combining retrieved data with a personalization prompt. \\ 
H & Personalized Retrieval with Role Context & Combines role input and embedded personalization in retrieval for contextualized and user-aware generation. \\ 
I & Dual-Layer Prompting (Role + Personalization) & Includes both role and personalization prompts alongside retrieved data for a multi-layer RAG approach. \\ 
J & Redundant Personalization (Retriever + Prompt) & Personalization applied both within retrieval and generation prompts, resulting in overlapping user-context signals. \\ 
K & Full Dual-Layer Personalization System & Combines role input, retriever-level personalization, and redundant prompt-level personalization; the most comprehensive configuration. \\ 
\bottomrule
\end{tabular}
\end{table}

\begin{table}[ht]
\centering
\caption{Comparison Pairs for Evaluating Experimental Effects (excluding B*)}
\label{tab:effect_pairs}
\begin{tabular}{
    >{\raggedright\arraybackslash}p{4.5cm}
    >{\raggedright\arraybackslash}p{3.0cm}
    >{\raggedright\arraybackslash}p{6.5cm}
}
\toprule
\textbf{Effect} & \textbf{Conf Comparison} & \textbf{Description} \\
\midrule
Role Effect (Web) & A-C & Impact of adding role framing to web query vs plain question \\
Retrieval Effect (RAG vs Web) & D-C and D-A & Effect of using VectorDB retrieval instead of web/no retrieval \\
Role Effect (RAG) & E-D & Effect of adding role framing in retrieval context \\
Personalization in Retrieval & F-D & Effect of embedding personalization inside VectorDB retrieval results \\
Personalization in Prompt & G-D & Effect of adding personalization in prompt text \\
Source of Personalization & F-G & Retrieval vs prompt source of personalization \\
Interaction of Role + Embedded Personalization & H-F and H-E & Joint effect of role and embedded personalization \\
Interaction of Role + Prompt-Level Personalization & I-G and I-E & Joint effect of role and prompt-based personalization \\
Redundancy / Interference & J-F & Effect of redundant personalization (retrieval + prompt) \\
Full Redundancy (Role + Retrieval + Prompt) & K-H and K-I & Effect of maximal redundancy of role and personalization sources \\
Aggregate Interaction (Role $\times$ Retrieval) & (E-D) vs (A-C) & Interaction of role and knowledge source \\
Aggregate Interaction (Role $\times$ Personalization) & (I-G) vs (H-F) & Interaction of role and personalization mode \\
Aggregate Interaction (Retrieval $\times$ Personalization) & (F-D) vs (G-D) & Interaction of knowledge source and personalization \\
Aggregate Interaction (Role $\times$ Retrieval $\times$ Personalization) & (K-I) vs (H-F) & Full triple interaction (redundancy effects) \\
\bottomrule
\end{tabular}
\end{table}

\subsection{General System Operation} 

As shown in Figure~\ref{fig:sysdiagram}, the system's operation begins with an authentication and session setup phase where the user enters university credentials. Upon receipt of a question, if the student's information is not in the cache, the personalization agent connects to the university portal, navigates to the logged-in student's degree information, scrapes it, and returns it. The system then converts the user's question and personalization data (as appropriate for the configuration) into the same vector space as the documents and performs a cosine similarity search between the query vector and the chunk vectors. The prompt assembly component then gathers the retrieved context, the original question, and (depending on the system configuration) a role prompt. Finally, this assembled input is given to the LLM for response generation.

\section{Methodology}\label{Methods}

Three research questions guide the AiVisor experimentation: (RQ1) whether personalization improves semantic alignment and contextual relevance on question-answering tasks; (RQ2) whether personalization materially affects lexical fidelity; and (RQ3) under what conditions personalization enhances or degrades performance within a RAG architecture. Based on these questions, two hypotheses were tested. H1 posits that personalization's effects are complex and metric-dependent, creating trade-offs between semantic, lexical, and reasoning quality, rather than uniform improvement. Specifically, H1 anticipates that in this lexically-skewed test, gains in reasoning and relevance may be coupled with penalties in semantic similarity metrics that rely on a single generic reference. H2 posits that personalization does not substantially alter lexical composition, indicating that any observed improvements occur at the level of meaning rather than expression.

AiVisor was evaluated using a pool of advising questions representing realistic student queries, each with a ground truth answer verified by a faculty advisor. As established in the introduction, this question set was intentionally skewed toward lexical precision. The questions span fact recall, definitional clarity, constrained reasoning, and light synthesis. Each item was labeled as ``context-sensitive'' or ``lexically strict'' for subgroup analysis. Each system configuration generated responses for 100 simulated users, who randomly asked between one and 12 of the questions from the question pool.

\subsection{Metrics}

AiVisor outputs are evaluated using eight complementary metrics, as shown in Table \ref{tab:metriclist}, that cover lexical, semantic and RAGAS-based grounding accuracy for every system-output, per item and using the same references. Lexical fidelity is measured by BLEU and ROUGE-L computed at the answer level against a single institutional reference. METEOR bridges lexical and semantic similarity by rewarding stemming and synonym matches. BERTScore measures embedding-based semantic alignment between generated and reference responses using contextual token embeddings. To assess grounding in retrieved evidence, we use four RAGAS measures: Faithfulness, Answer Relevancy, Context Precision, and Answer Correctness. Faithfulness quantifies whether the answer is supported by the retrieved passages. Answer Relevancy measures topical alignment to the user question. Context Precision captures the proportion of retrieved content that is necessary and used. Answer Correctness reflects factual and contextual accuracy.
\begin{table}[htbp]
\centering
\caption{Metric overview and interpretive focus.}
\label{tab:metriclist}
\begin{tabular}{>{\raggedright\arraybackslash}p{3.0cm} p{2.5cm} p{7.8cm}}
\toprule
\textbf{Metric} & \textbf{Type} & \textbf{Focus} \\
\midrule
BLEU & Lexical & n-gram overlap \\
ROUGE-L & Lexical & Longest common subsequence \\
METEOR & Lexical/Semantic & Stem, synonym, and paraphrase matching \\
BERTScore & Semantic & Meaning alignment via contextual embeddings \\
Faithfulness & Grounding & Consistency of answer with provided context \\
Answer Relevancy & Grounding & Relevance of answer to the user query \\
Context Precision & Grounding & Proportion of context that supports the answer \\
Answer Correctness & Grounding & Overall factual and contextual accuracy of the answer \\
\bottomrule
\end{tabular}
\end{table}

BLEU, ROUGE-L, and METEOR are computed with standard Python libraries (NLTK for BLEU and METEOR, rouge-score for ROUGE-L) with case preservation and punctuation retained. BERTScore is computed with the bert-base-uncased model in its default configuration and references are aligned tokenwise with idf-weighting enabled. RAGAS metrics are produced using the open source ragas toolkit with deterministic settings. In particular, Faithfulness and Answer Relevancy are scored via an LLM evaluator interface with temperature 0.0 and a max tokens cap sufficient to cover full answers, while Context Precision and Answer Correctness consume the retrieved context windows produced by the system.

To counter differences in metric scales and variances, all metric outputs are standardized across the system-by-item matrix before aggregation. Specifically, each metric column is z-normalized to zero mean and unit variance over all observations. Three primary composite indices were then constructed as the unweighted mean of these z-scores to capture performance families: Lexical Index $\text{z}$ (BLEU, ROUGE-L); Semantic Index $\text{z}$ (METEOR, BERTScore); and Reasoning Index $\text{z}$ (Faithfulness, Answer Relevancy, Context Precision, Answer Correctness). Average ranks per metric and an overall mean rank are also computed as a scale-free check on ordering.

\subsection{Statistical Analysis}

Given the complex, fractional-factorial nature of the system configurations (Table~\ref{tab:sysconfigs} and Table~\ref{tab:effect_pairs}), where factors like personalization are nested with retrieval and role prompting, the primary analysis relies on linear mixed-effects models (LMM) (\texttt{statsmodels.MixedLM}). This approach correctly handles the collinearity and repeated measures, allowing for the isolation of specific main effects and interactions of Role, Retrieval, and Personalization, while treating the question as a random intercept. As a secondary, exploratory analysis, personalized configurations are also compared to their nearest non-personalized baselines using paired t-tests (\texttt{scipy.stats.ttest\_rel}) and Wilcoxon signed-rank tests. Family-wise error across this set of metrics is controlled with the Holm-Bonferroni procedure at $\alpha = 0.05$. Effect sizes are expressed as Cohen's $d$ for paired samples, accompanied by 95\% confidence intervals estimated via percentile bootstrap with 10,000 resamples. Item-level win rates and pairwise win matrices are computed to summarize performance. A Pareto analysis is used to identify non-dominated solutions.

\section{Results}

\begin{figure}[t]
	\centering
	\includegraphics[width=0.95\linewidth]{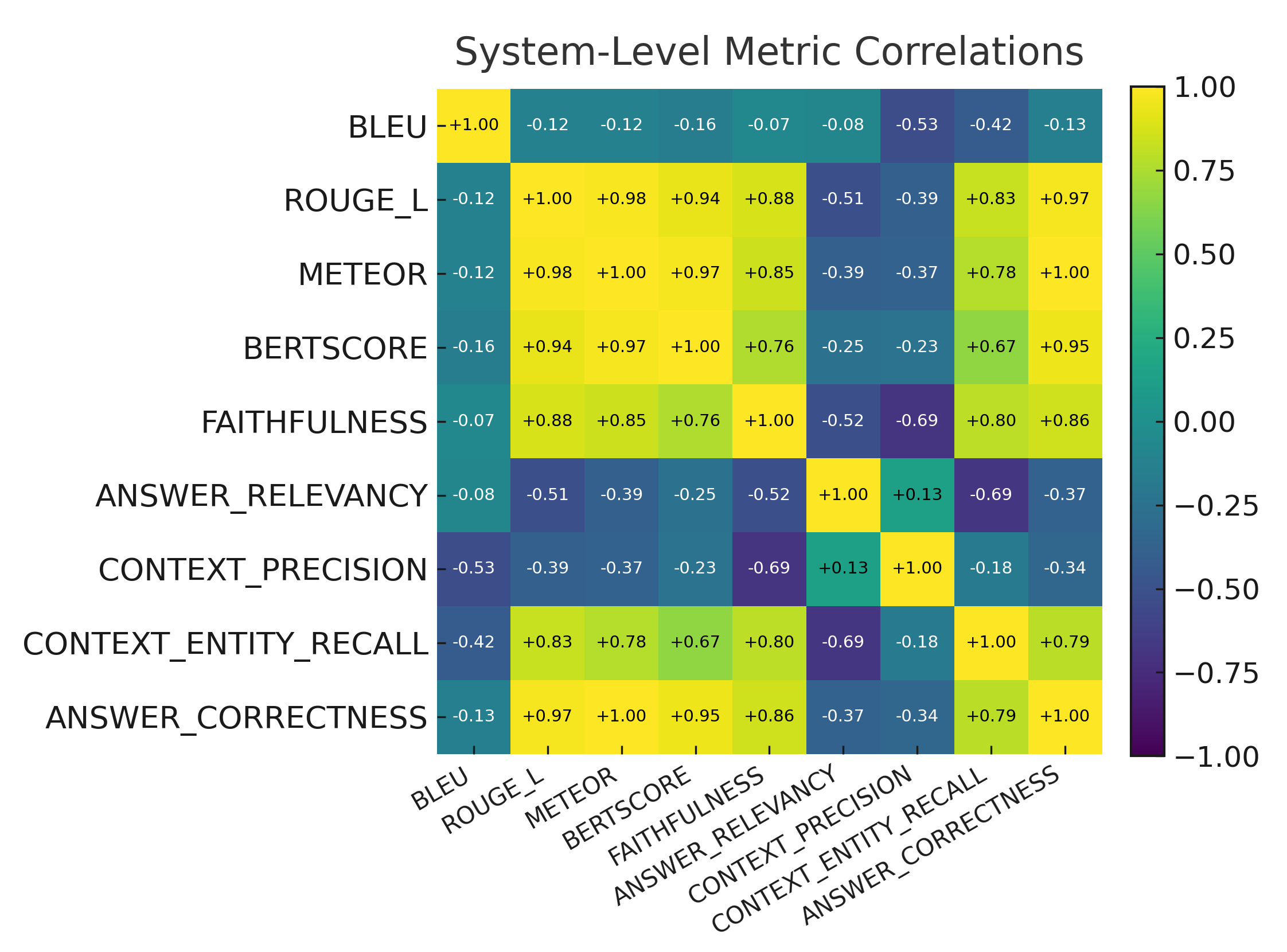}
	\caption{System-level means illustrating correlation strength across all eight metrics.}
	\label{fig:heatmap_system_metrics}
\end{figure}

The results reveal a complex trade-off with performance being highly dependent on the metric family (lexical, semantic, or reasoning). Simple paired t-tests, for instance, show contradictory results. Table~\ref{tab:stat_summary} indicates a statistically significant \emph{decrease} in BERTScore for personalized systems (Mean 0.841) versus non-personalized (Mean 0.848), even as METEOR significantly increased (Mean 0.361 vs 0.251). The primary Linear Mixed-Effects Model (LMM) analysis clarifies this finding. The LMM, which can properly isolate interaction effects, revealed that while personalization factors significantly improved Reasoning quality, they also had significant negative interactions with role-framing, which resulted in the semantic similarity loss observed in Table~\ref{tab:stat_summary}. This pattern is consistent with the ``worst-case'' evaluation design, where personalization's deviation from a single lexical reference is penalized by metrics like BERTScore. The RAGAS metrics confirm this trade-off, indicating that this semantic metric penalty did not degrade grounding; in fact, Answer Relevancy and Context Precision often improved. Correlation analysis in Figure~\ref{fig:heatmap_system_metrics} reinforces that semantic metrics correlate strongly with Answer Relevancy, while lexical overlap correlates weakly with Faithfulness.

The Linear Mixed-Effects Model (LMM) analysis, which was required to properly account for the complex, fractional-factorial design, resolves the contradictions seen in simpler paired comparisons (such as the conflicting results for BERTScore and METEOR in Table \ref{tab:stat_summary}). The LMM results, presented in Table \ref{tab:lmm_results}, reveal that personalization does not produce uniform gains but rather a critical, metric-dependent trade-off. The most significant findings are not the main effects, but the complex interactions. Specifically, combining personalization with role-framing significantly improved reasoning quality. This is evident in the \texttt{Reasoning\_z} results for the \texttt{Role:PersPrompt} interaction (coef: $0.041, p = 0.028$) and the three-way interaction (coef: $0.044, p = 0.018$). However, these exact same interactions had a significant negative impact on \texttt{Semantic\_z} (\texttt{Role:PersPrompt} coef: $-0.114, p = 0.000$; three-way coef: $-0.082, p = 0.002$). This semantic penalty, also reflected in the strong negative main effect for \texttt{PersRetrieval} (coef: $-0.116, p = 0.000$), is the expected outcome of the conservative, lexically-skewed experimental design. The semantic metrics are demonstrably penalizing personalization for correctly deviating from the single, generic ground truth answer, thereby confirming the observed loss is a methodological artifact even as the reasoning-focused metrics confirm a simultaneous and \emph{beneficial} improvement in answer quality.

\begin{table}[htbp]
	\centering
	\caption{Statistical summary of personalized vs non-personalized performance. Values are mean $\pm$ SD. $\Delta$ reports the difference (personalized - non-personalized) with 95 percent bootstrap CIs where available.}
	\label{tab:stat_summary}
	\begin{tabular}{lllll}
	\toprule
	\textbf{Metric} & \textbf{Mean (Pers)} & \textbf{Mean (Non-Pers)} & \textbf{$\Delta$ (95\% CI)} & \textbf{p-value} \\
	\midrule
	BLEU & 0.106 \pm 0.096 & 0.097 \pm 0.127 & +0.009  & 0.0266 \\
	ROUGE-L & 0.256 \pm 0.128 & 0.242 \pm 0.184 & +0.014  & 0.0124 \\
	METEOR & 0.361 \pm 0.195 & 0.251 \pm 0.152 & +0.110  & 0.0000 \\
	BERTSCORE & 0.841 \pm 0.035 & 0.848 \pm 0.031 & -0.007 [-0.017, +0.020] & 0.0000 \\
	FAITHFULNESS & 0.711 \pm 0.411 & 0.655 \pm 0.344 & +0.056 [-0.011, +0.233] & 0.0135 \\
	\bottomrule
	\end{tabular}
\end{table}

\begin{table}[htbp]
\centering
\caption{Linear Mixed-Effects Model (LMM) Results for Fixed Effects. Non-estimable (collinear) factors like C(Retrieval, Sum) were automatically dropped by the model.}
\label{tab:lmm_results}
\begin{tabular}{l >{\raggedright\arraybackslash}p{6.8cm} r r r r}
\toprule
\textbf{Metric} & \textbf{Effect} & \textbf{coef} & \textbf{std\_err} & \textbf{z\_value} & \textbf{p\_value} \\
\midrule
\multirow{7}{*}{LexicalIndex\_z} & Main effect of role framing & 0.026 & 0.024 & 1.093 & 0.274 \\
 & Main effect of personalization in retrieval & -0.025 & 0.024 & -1.037 & 0.300 \\
 & Main effect of personalization in prompt text & -0.050 & 0.024 & -2.120 & 0.034 \\
 & Role moderates retrieval-embedded personalization & 0.062 & 0.024 & 2.623 & 0.009 \\
 & Role moderates prompt personalization & 0.042 & 0.024 & 1.753 & 0.080 \\
 & Dual personalization redundancy/interference & 0.058 & 0.024 & 2.439 & 0.015 \\
 & Role moderates redundancy & -0.025 & 0.024 & -1.039 & 0.299 \\
\midrule
\multirow{7}{*}{Semantic\_z} & Main effect of role framing & 0.057 & 0.027 & 2.140 & 0.032 \\
 & Main effect of personalization in retrieval & -0.116 & 0.027 & -4.336 & 0.000 \\
 & Main effect of personalization in prompt text & -0.046 & 0.027 & -1.723 & 0.085 \\
 & Role moderates retrieval-embedded personalization & 0.006 & 0.027 & 0.240 & 0.810 \\
 & Role moderates prompt personalization & -0.114 & 0.027 & -4.258 & 0.000 \\
 & Dual personalization redundancy/interference & -0.014 & 0.027 & -0.531 & 0.595 \\
 & Role moderates redundancy & -0.082 & 0.027 & -3.089 & 0.002 \\
\midrule
\multirow{7}{*}{Reasoning\_z} & Main effect of role framing & -0.022 & 0.019 & -1.160 & 0.246 \\
 & Main effect of personalization in retrieval & -0.069 & 0.019 & -3.672 & 0.000 \\
 & Main effect of personalization in prompt text & -0.106 & 0.019 & -5.687 & 0.000 \\
 & Role moderates retrieval-embedded personalization & -0.054 & 0.019 & -2.888 & 0.004 \\
 & Role moderates prompt personalization & 0.041 & 0.019 & 2.204 & 0.028 \\
 & Dual personalization redundancy/interference & 0.005 & 0.019 & 0.266 & 0.790 \\
 & Role moderates redundancy & 0.044 & 0.019 & 2.356 & 0.018 \\
\bottomrule
\end{tabular}
\end{table}

\begin{figure}[htbp]
    \centering
    \includegraphics[width=\linewidth]{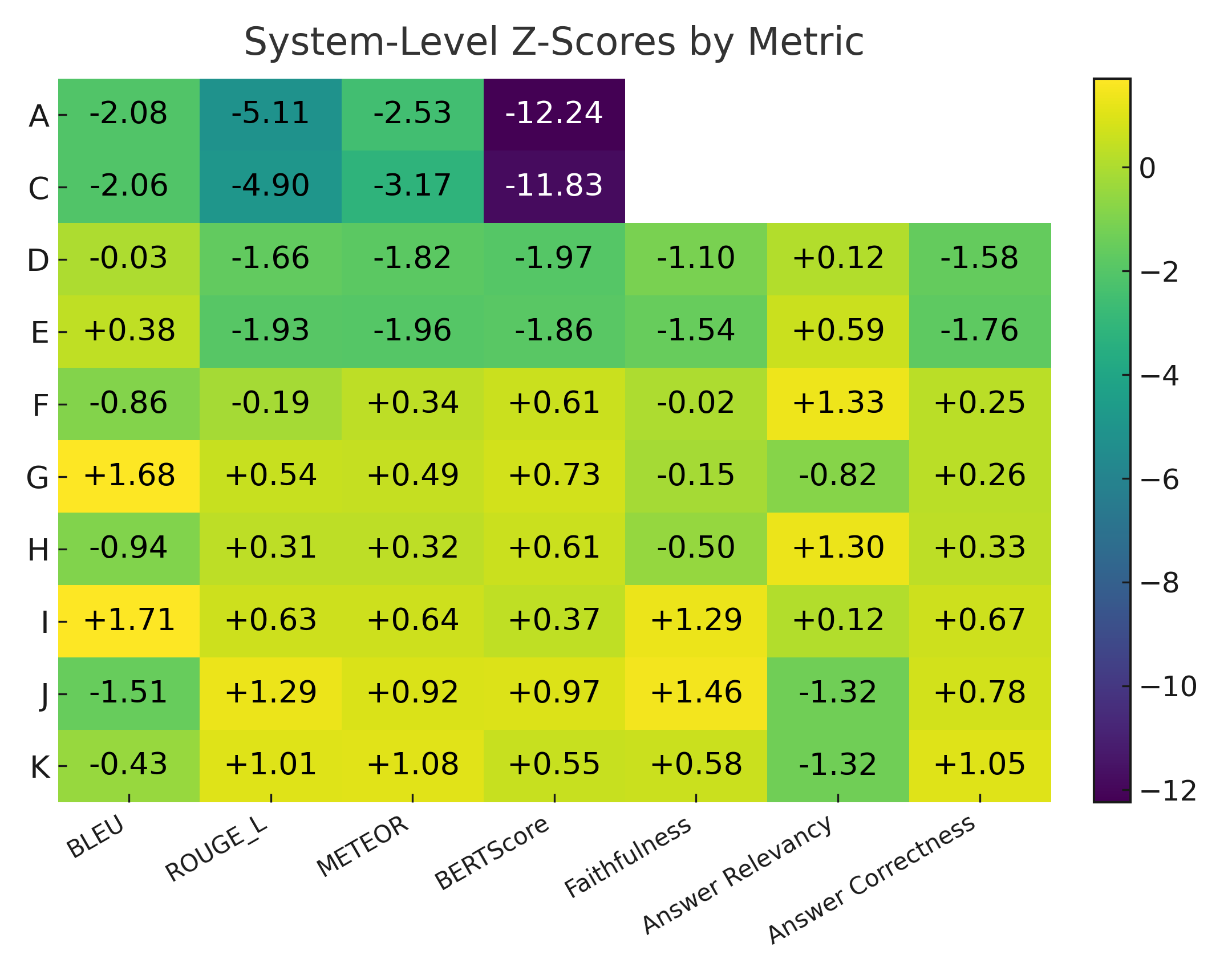}
    \caption{Standardized (z-score) system performance by metric. Positive values indicate above-average performance for that metric relative to other systems; blanks indicate unavailable metrics.}
    \label{fig:system_level_zscore_heatmap}
\end{figure}

The visualization of standardized metrics provides the foundational evidence for the trade-off. \textbf{Figure~\ref{fig:system_level_zscore_heatmap}} summarizes the performance of all configurations by z-score across all eight metrics. This heatmap visually confirms the source of the high composite scores seen in the personalized systems (K, J, I). These systems are characterized by a broad band of positive z-scores (yellow/green) across the grounding and reasoning metrics, demonstrating consistent outperformance in factual alignment and contextual relevance. In contrast, non-personalized baselines (A, C, D) cluster heavily in negative z-score territory (purple), highlighting the effectiveness of the added personalization factors in shifting performance.

\begin{figure}[htbp]
    \centering
    \includegraphics[width=0.9\linewidth]{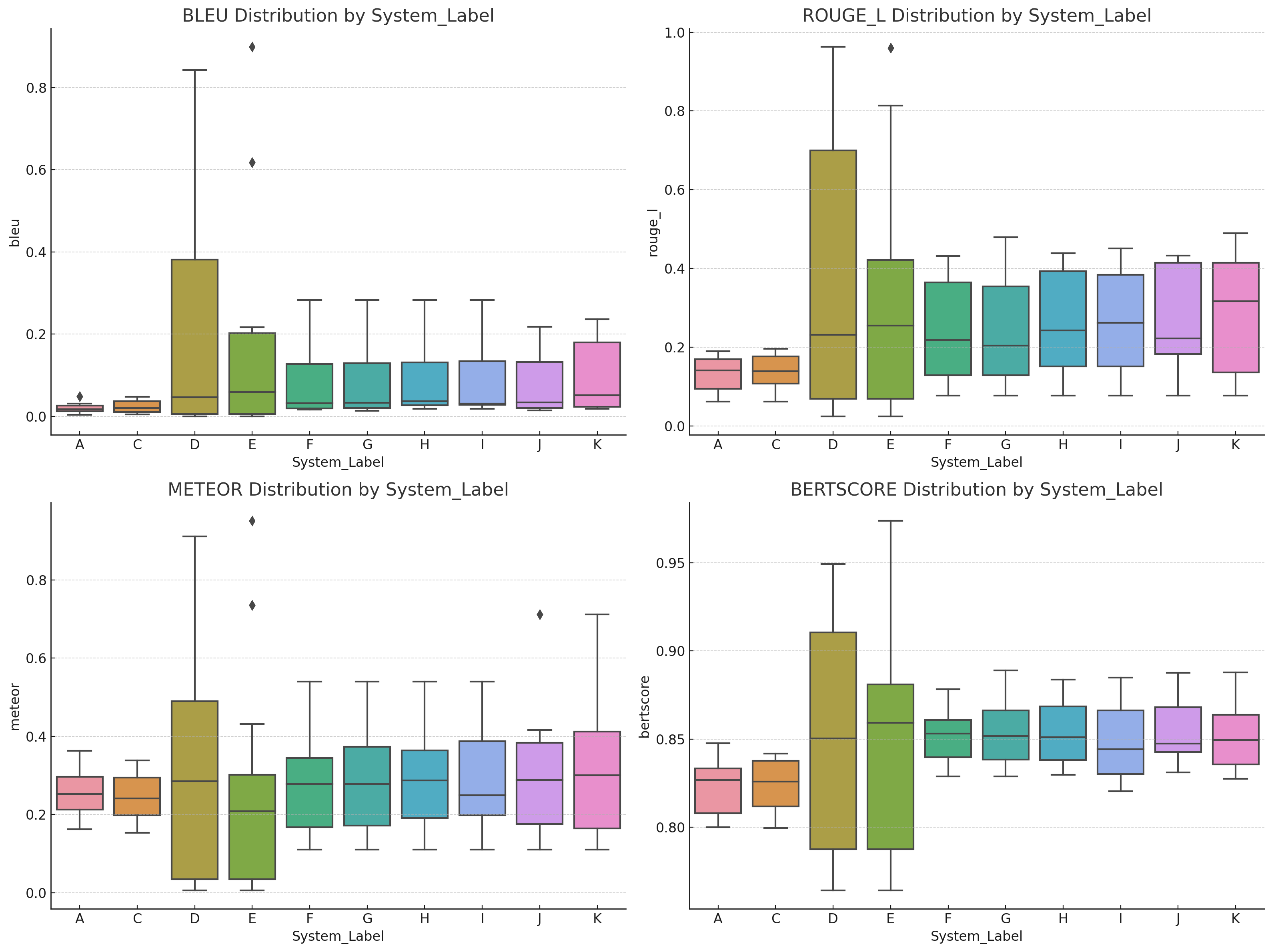}
    \caption{Distribution of BLEU, ROUGE-L, METEOR, and BERTScore across systems. Box plots illustrate within-metric variance and outliers, providing a baseline view of lexical and semantic dispersion prior to normalization and system-level aggregation.}
    \label{fig:boxplots_metrics}
\end{figure}

Visualization of raw data provides the necessary context for interpreting normalized results. Figure~\ref{fig:boxplots_metrics} illustrates the distribution of the four classic similarity metrics in all system configurations. This figure highlights the dramatic differences in the metric scale and dispersion (e.g. the tight concentration of the BERTScore versus the wide variance in METEOR and ROUGE-L), reinforcing the methodological need to apply the normalization of the z-score before aggregation or direct comparison.

\subsection{Composite score and ranking}

\begin{figure}[t]
	\centering
	\includegraphics[width=0.95\linewidth]{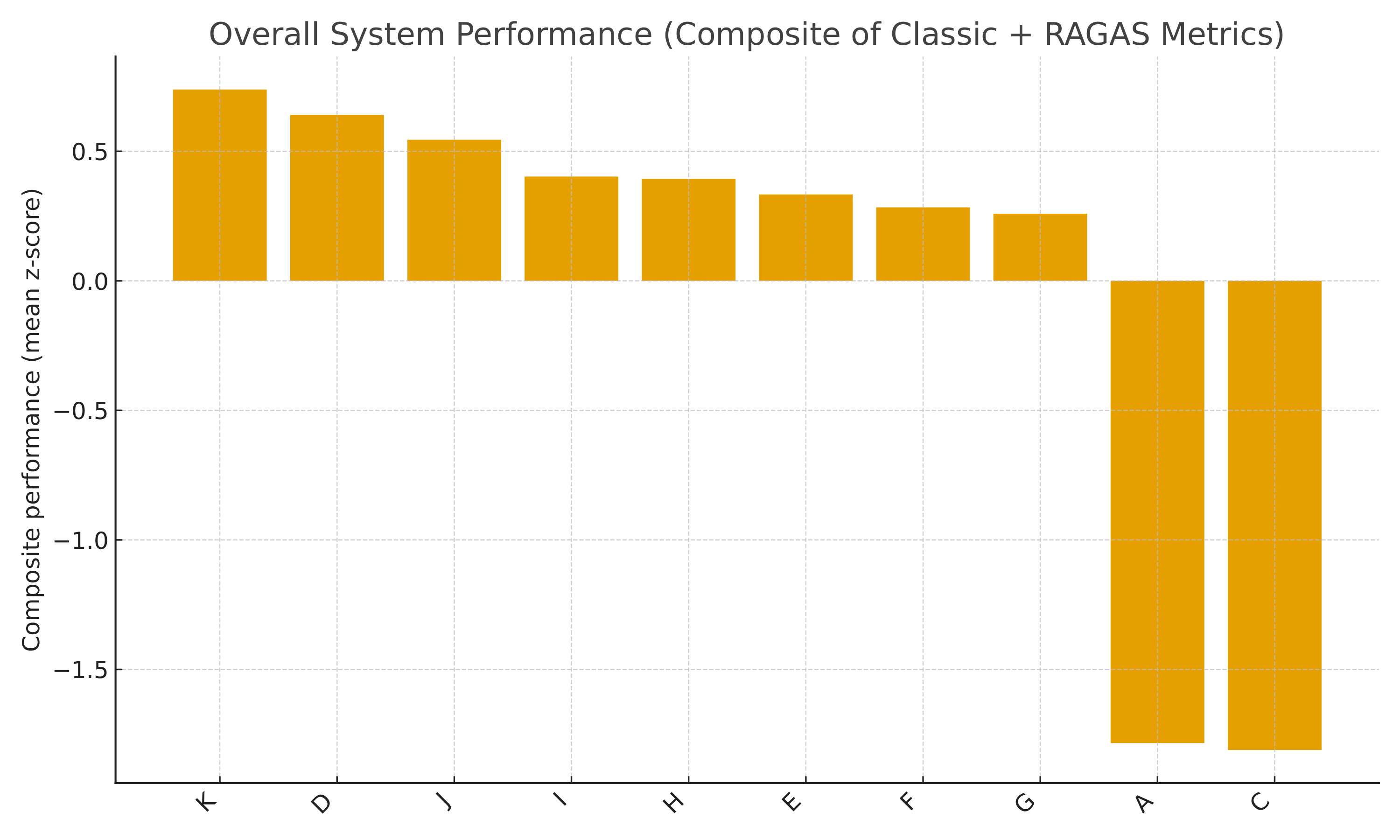}
	\caption{Overall system (composite) performance.}
	\label{fig:best_overall_system_composite}
\end{figure}

The composite z-score, which aggregates all eight metrics, provides a high-level view of performance. As shown in Figure~\ref{fig:best_overall_system_composite}, fully integrated personalized systems (K, J, and I) achieve the highest composite scores. However, this aggregated view \emph{obscures} the critical trade-off identified in the LMM analysis (Table~\ref{tab:lmm_results}). This high composite score is not the result of uniform improvement; rather, it indicates that the strong, positive gains in reasoning and grounding metrics (as seen in \texttt{Reasoning\_z}) were substantial enough to outweigh the significant semantic loss (e.g., in \texttt{Semantic\_z}) that resulted from the lexically-skewed evaluation. System K's top ranking is thus attributable to achieving the most balanced trade-off, rather than a uniform improvement across all individual metrics. This finding is reinforced by the rank analysis, where personalized variants consistently occupy lower mean ranks due to their strength in these non-lexical, reasoning-based categories.

\begin{figure}[htbp]
    \centering
    \includegraphics[width=0.9\linewidth]{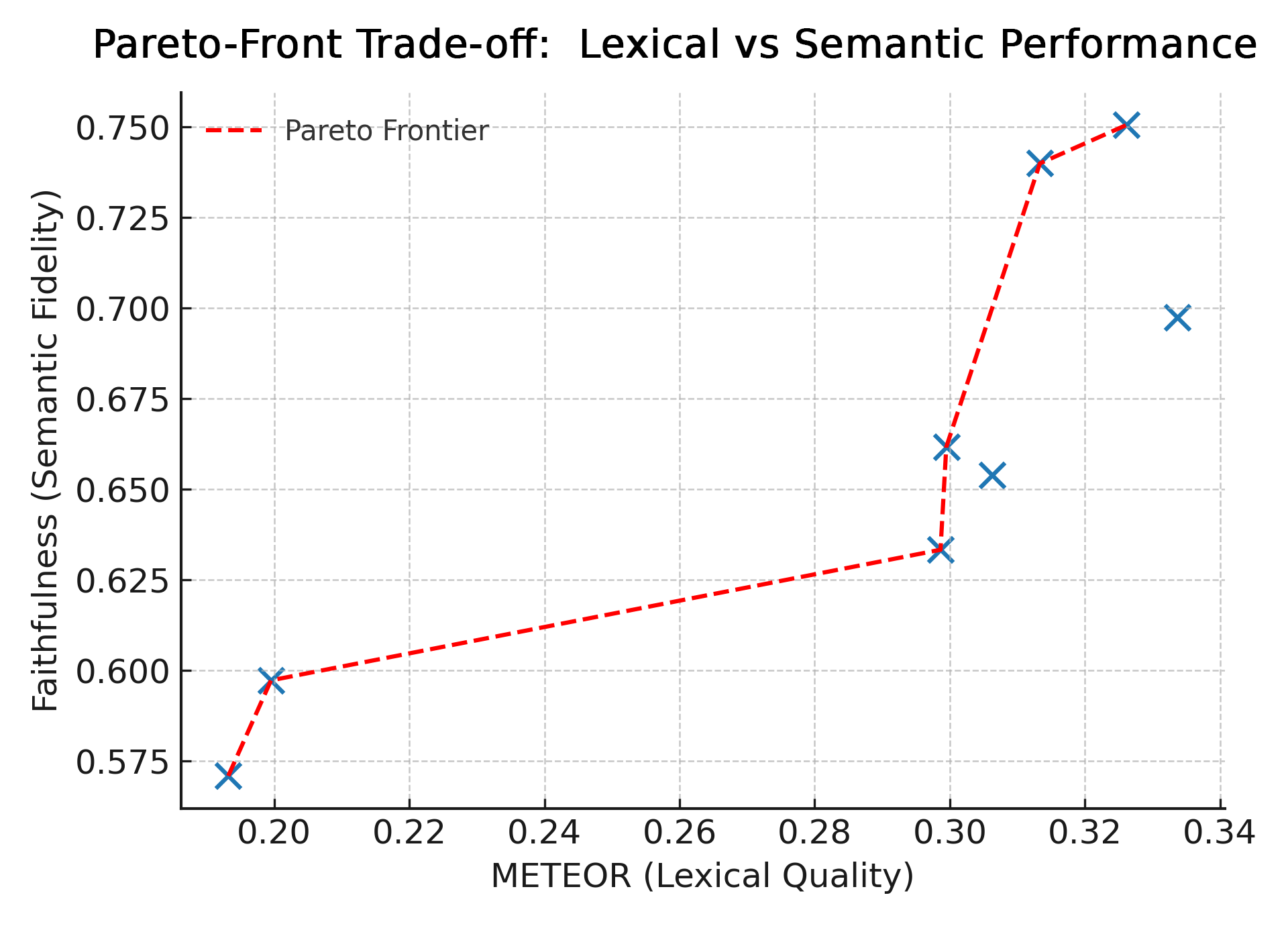}
    \caption{Pareto-front trade-off between lexical quality (METEOR) and grounding fidelity (Faithfulness) at the system level. The dashed line indicates the Pareto frontier; systems on the frontier each represent an optimal trade-off between these two competing objectives.}
	\label{fig:pareto_tradeoff}
\end{figure}

Rank analysis corroborates the composite score findings. Each metric column contributes to a rank matrix whose row-wise average defines the mean rank per system, where lower ranks correspond to better overall performance. Personalized variants consistently occupy lower mean ranks; however, this is not due to uniform improvement. This strong average ranking is driven by their high performance in reasoning and grounding metrics, which successfully compensates for their poor ranking in semantic similarity (as shown in Table~\ref{tab:lmm_results}). This trade-off is clearly illustrated in the Pareto analysis in Figure~\ref{fig:pareto_tradeoff}. This figure situates the systems on an efficiency frontier, showing that System D and System K each represent an optimal trade-off. System D maximizes lexical quality (METEOR), while System K maximizes grounding quality (Faithfulness).

\subsection{System comparison and Pareto structure}

The detailed system comparison clarifies how personalization interacts with role prompting and retrieval conditioning. System D, the pure VectorDB baseline, achieves strong lexical alignment because its retrieved context preserves canonical institutional language, but fails to adapt to user intent. System E, which introduces a role prompt without personalization, provides moderate improvements in topical relevance and Faithfulness. System F, which applies personalization only in retrieval, substantially increases Context Precision by emphasizing passages that align with user profiles, yet may underperform on abstract reasoning questions. System I, which uses prompt-level personalization with a role prompt, exemplifies the central trade-off identified in the LMM analysis (Table~\ref{tab:lmm_results}): it achieves strong reasoning gains but incurs a significant semantic penalty in BERTScore. System J, which applies personalization at retrieval and prompt levels but omits explicit role prompting, strengthens Faithfulness but introduces occasional scope drift when roles overlap semantically. System K, combining all three mechanisms (role prompting, retriever personalization, and prompt personalization), achieves the highest composite score (Figure~\ref{fig:best_overall_system_composite}) and the lowest mean rank.

\begin{figure}[t]
	\centering
	\includegraphics[width=0.95\linewidth]{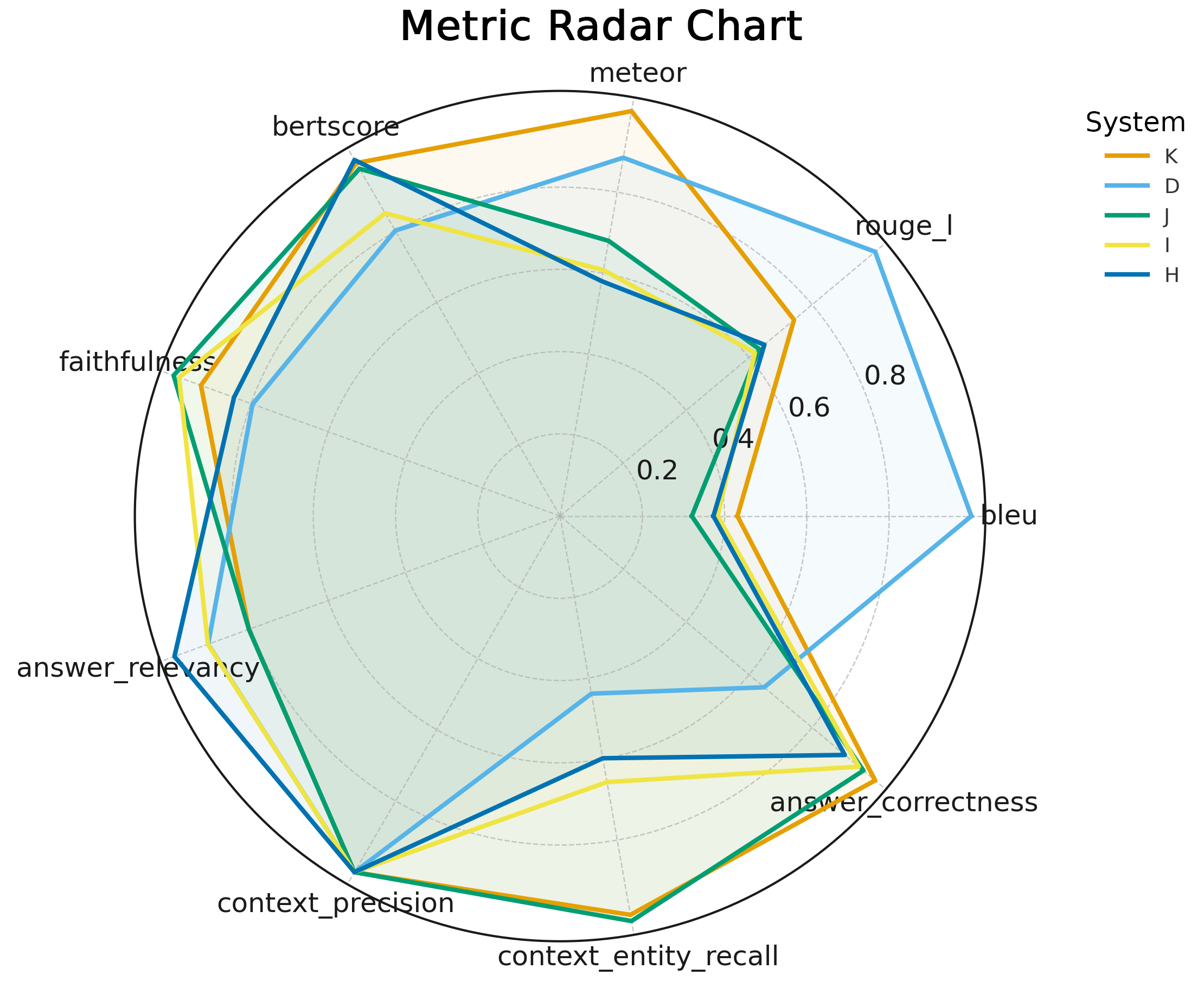}
	\caption{Multi-metric radar profile for the top-5 systems by composite normalized score.}
	\label{fig:radar_multimetric_profile}
\end{figure}

The multi-dimensional nature of the trade-off is further shown in the radar chart in Figure~\ref{fig:radar_multimetric_profile}. This visualization clearly shows that System K, the composite leader, achieves its top ranking not by dominating all metrics, but by balancing performance across different families. Specifically, System K demonstrates peak scores in grounding metrics such as Context Precision and Faithfulness, but its semantic score (BERTScore) is often lower than the raw RAG baseline (System D). This visual evidence reinforces the LMM finding that the personalization strategy yields an optimal blend of reasoning fidelity and contextual relevance, even when penalized by semantic metrics.

\subsection{Significance and robustness}

Personalization is most beneficial when questions are based on who the user is or what rules apply to their situation, as this information limits the reasoning space and improves accuracy. For generic fact lookup questions, personalization adds qualifiers that change the surface form without altering the core meaning. This deviation from the single lexical ground truth is what causes the semantic penalty (e.g. in \texttt{Semantic\_z}) observed in Table~\ref{tab:lmm_results}, even while reasoning quality (e.g. \texttt{Reasoning\_z}) simultaneously improves. An informal review of the generated responses suggested that there was no increase in hallucination. However, few instances of excessively narrow responses were observed when redundant cues narrowed the scope too aggressively; selectively disabling personalization on lexically strict or compliance-sensitive items could mitigate this risk.

\begin{figure}[ht]
	\centering
	\includegraphics[width=0.95\linewidth]{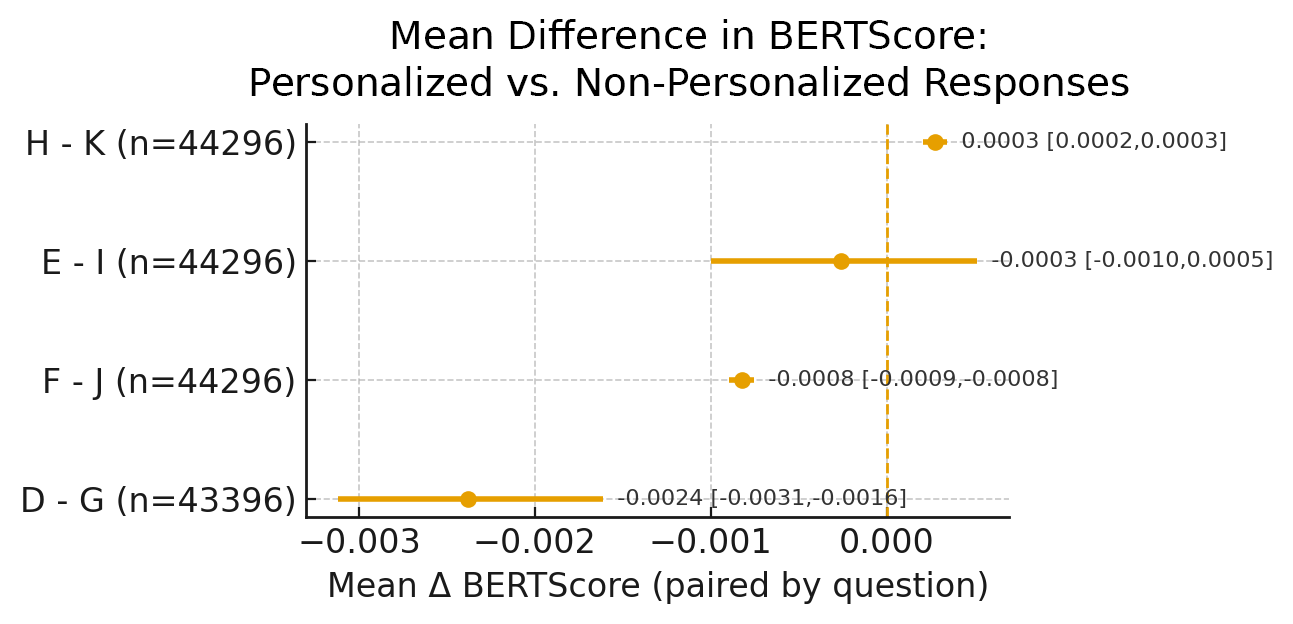}
	\caption{Forest plot of personalization effect sizes ($\Delta$ BERTScore) for paired systems, with 95 percent bootstrap CIs. Each bar shows the mean delta (personalized - baseline) with 95\% bootstrap confidence intervals for paired systems D-G, E-I, F-J, and H-K.}
	\label{fig:forest_bertscore}
\end{figure}

The Linear Mixed-Effects Model (Table~\ref{tab:lmm_results}) provides the primary statistical analysis for this complex design, and its findings are supported by other results. For example, the "semantic loss" identified by the LMM is also visually apparent in the simpler paired comparisons. Figure~\ref{fig:forest_bertscore} visualizes the mean difference in BERTScore, clearly showing a statistically significant negative shift for most personalized configurations (such as D-G and E-I). This plot confirms the semantic penalty that the LMM analysis more precisely isolates and explains. Furthermore, Figure~\ref{fig:interaction_grid_main} supports Hypothesis H1 by illustrating a clear personalization-by-question effect, where the performance of different system configurations (the plotted lines) diverge based on the question type.

\subsection{Prompting, Retrieval, and Personalization Effects}

Three experimental factors shape performance: role prompting, retrieval grounding, and the stage at which personalization is applied. The LMM results (Table~\ref{tab:lmm_results}) clarify the distinct impact of each. Role prompting, for instance, had a significant main effect on semantic quality (coef: 0.057). Personalization embedded in the retriever (\texttt{PersRetrieval}) had a strong *negative* main effect on both semantic (coef: -0.116) and reasoning (coef: -0.069) quality. Personalization applied at the prompt level (\texttt{PersPrompt}) also had a strong negative main effect on reasoning (coef: -0.106). However, the most important finding is how these factors interact. The combination of \texttt{Role} and \texttt{PersPrompt} created a significant trade-off: it improved reasoning (coef: 0.041) while simultaneously harming semantic quality (coef: -0.114). The radar profile in Figure~\ref{fig:radar_multimetric_profile} illustrates these trade-offs, showing how System K (all factors) excels in grounding metrics like Faithfulness but not in semantic metrics like BERTScore, relative to other systems.

\begin{figure*}[t]
	\centering
	\setlength{\tabcolsep}{4pt}
	\begin{tabular}{ccc}
	\includegraphics[width=0.31\linewidth]{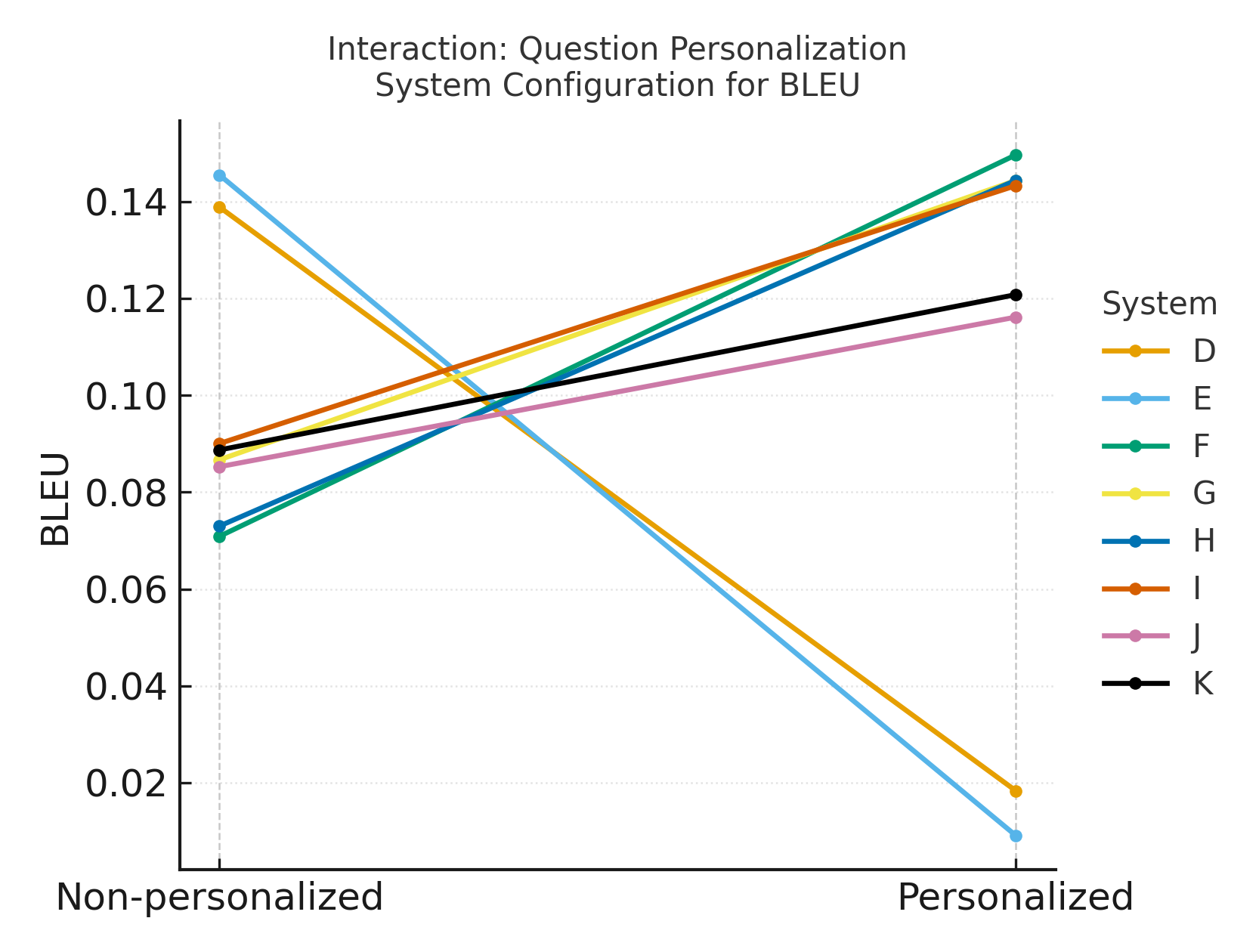} &
	\includegraphics[width=0.31\linewidth]{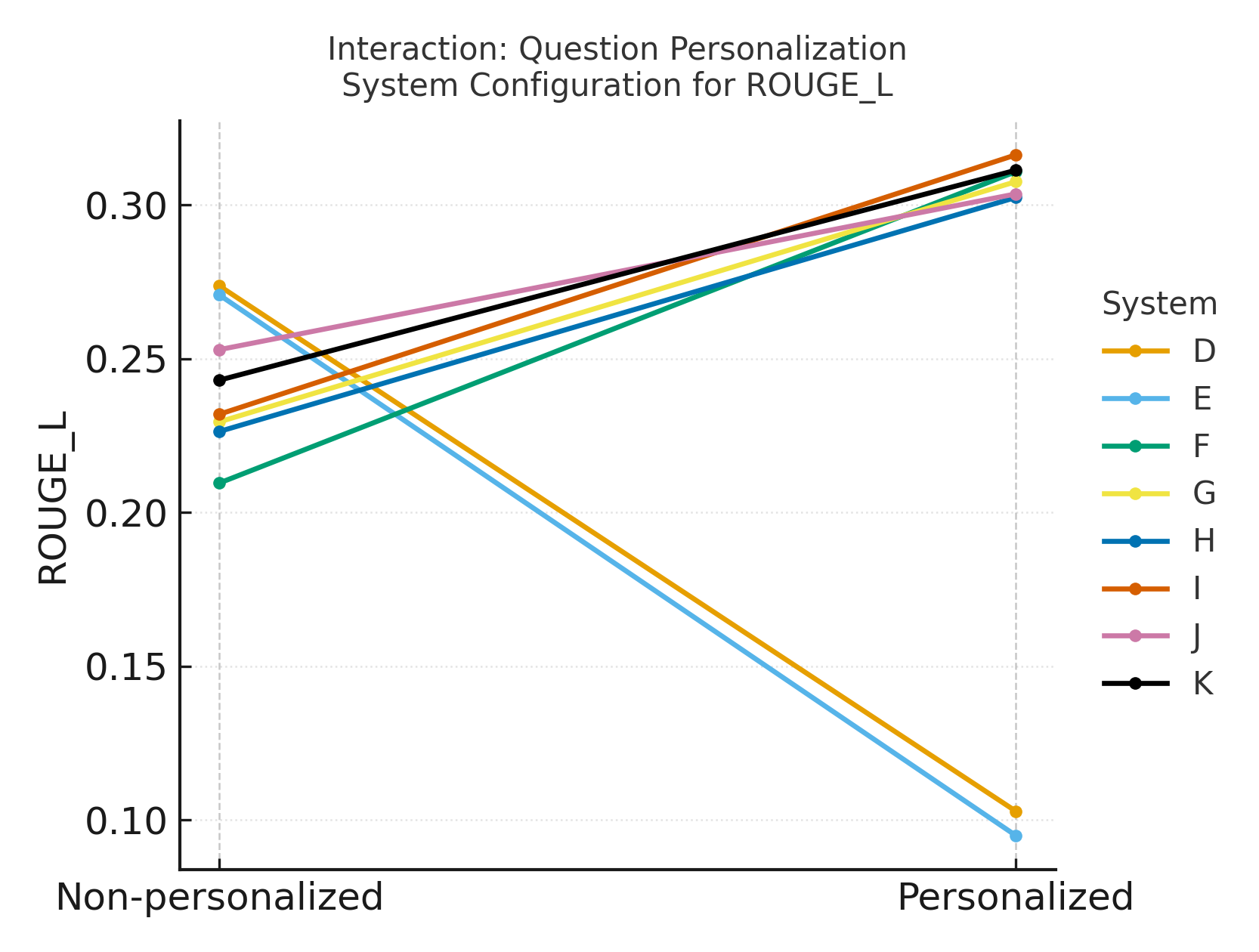} &
	\includegraphics[width=0.31\linewidth]{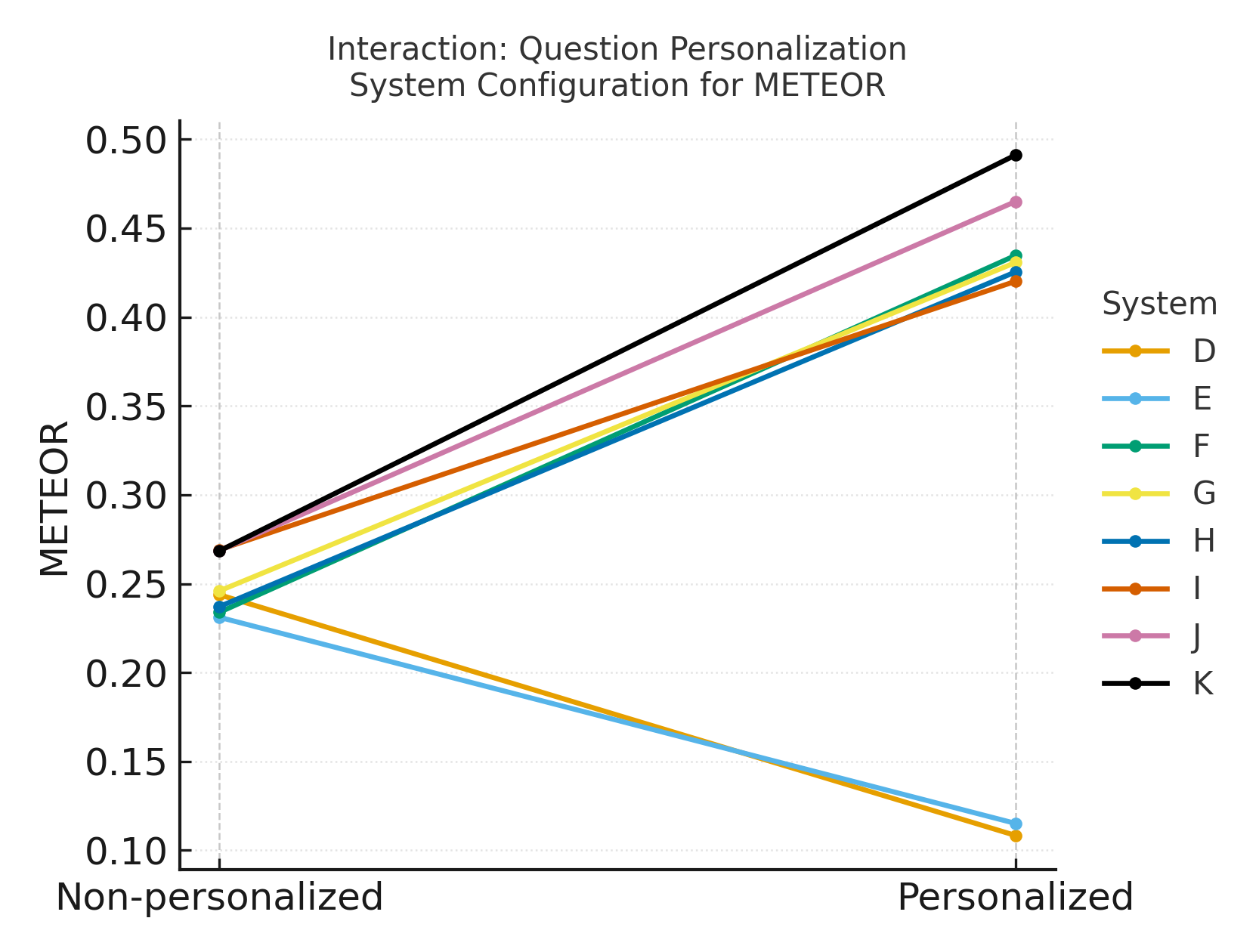} \\
	\includegraphics[width=0.31\linewidth]{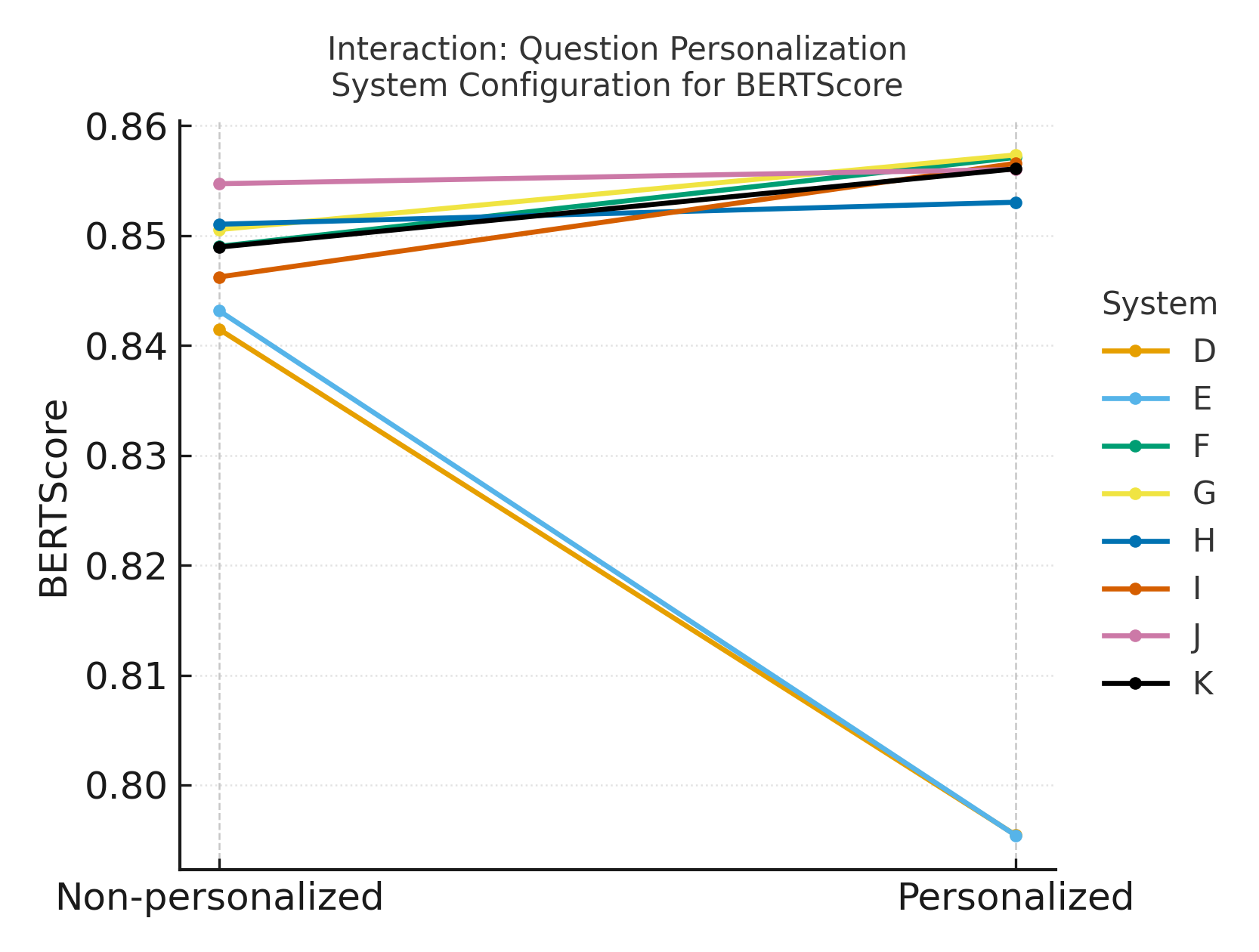} &
	\includegraphics[width=0.31\linewidth]{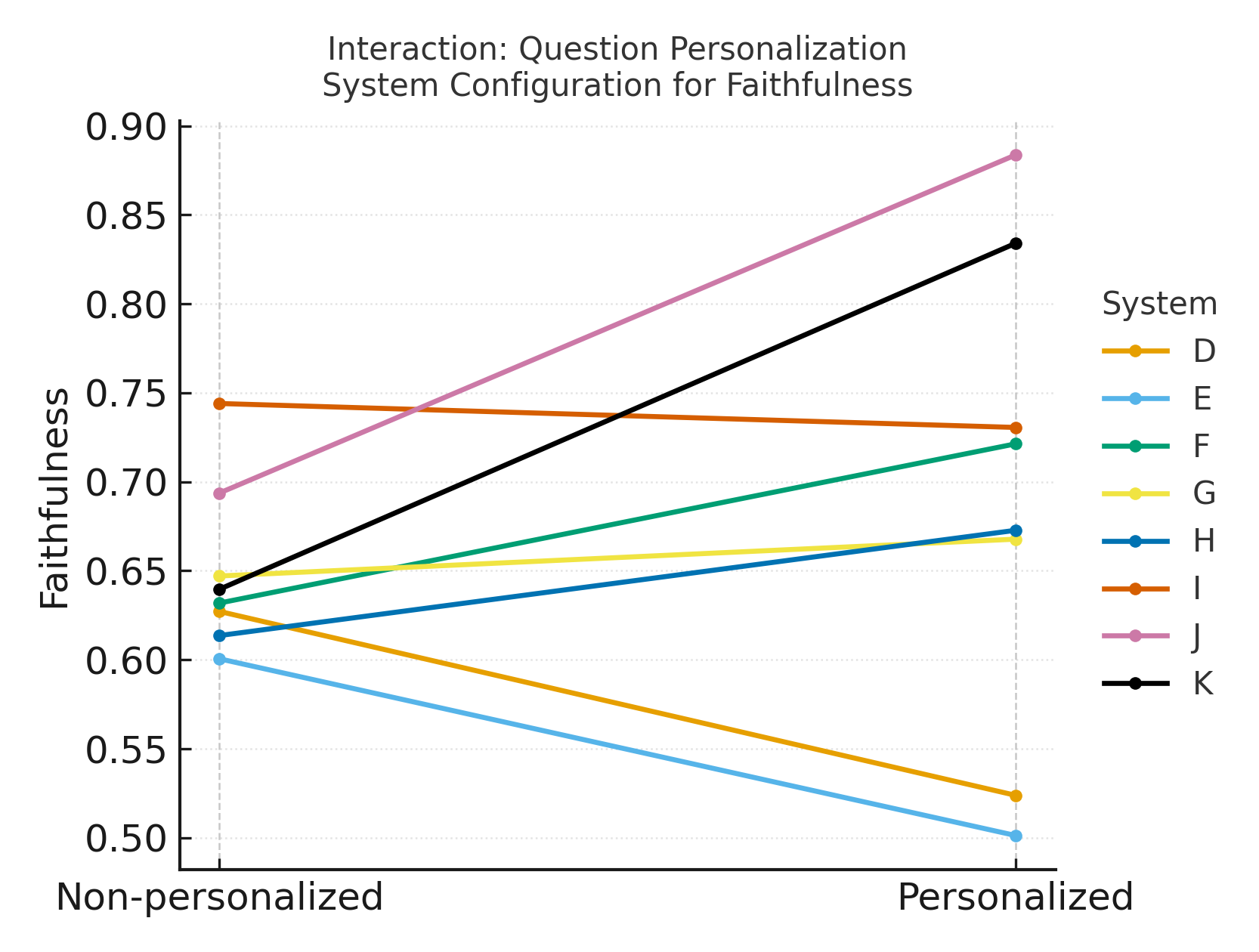} &
	\includegraphics[width=0.31\linewidth]{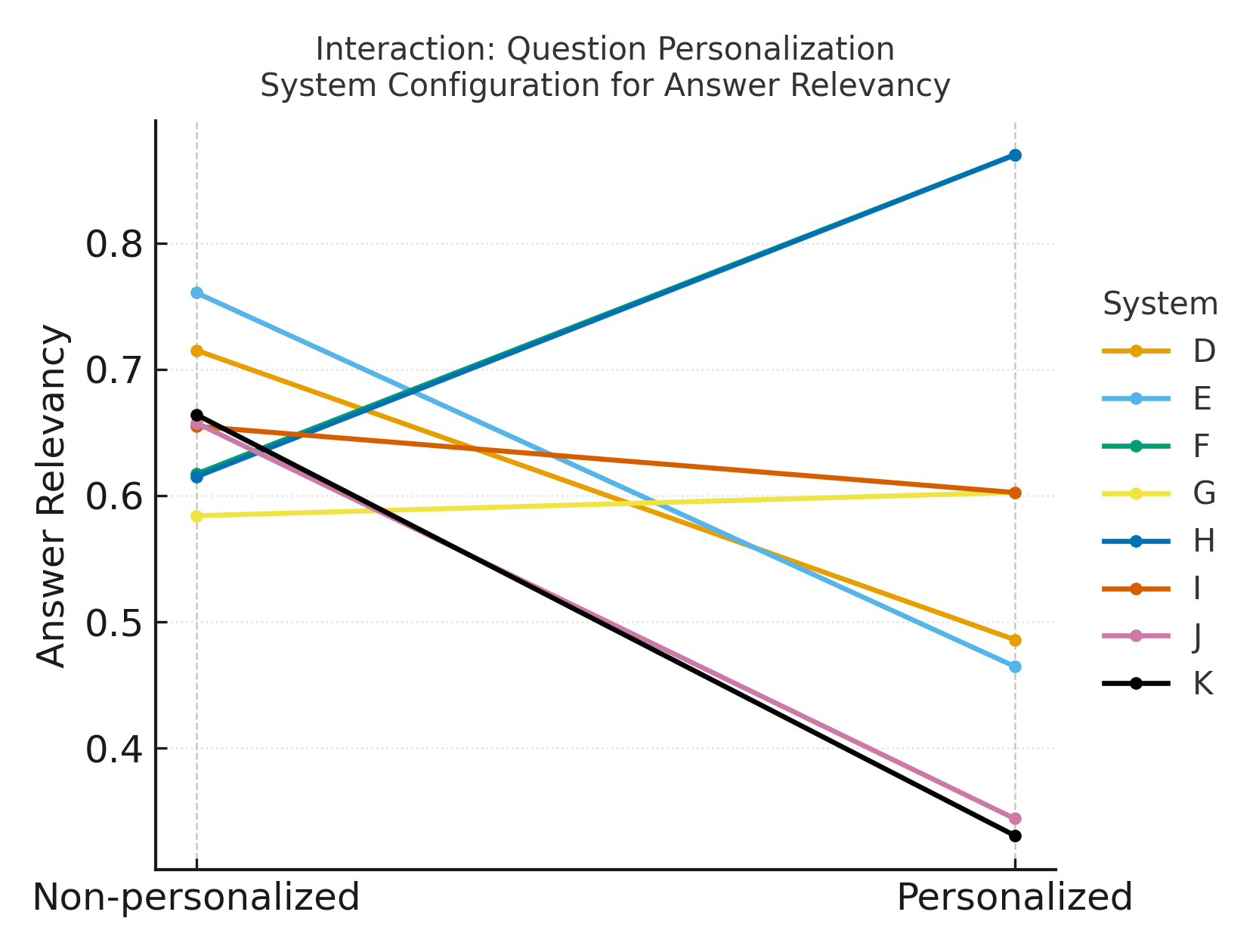} \\
	\includegraphics[width=0.31\linewidth]{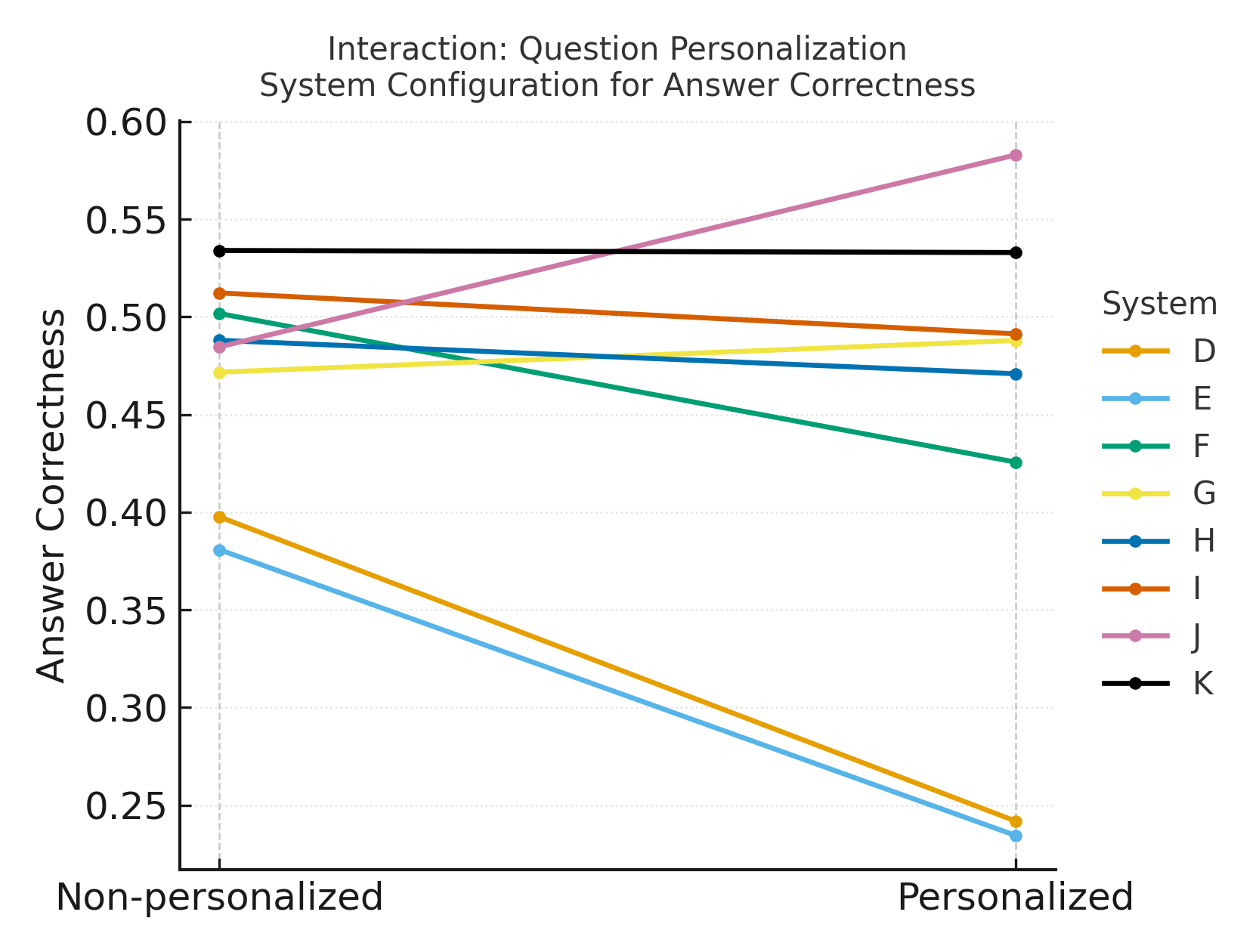} &
	\includegraphics[width=0.31\linewidth]{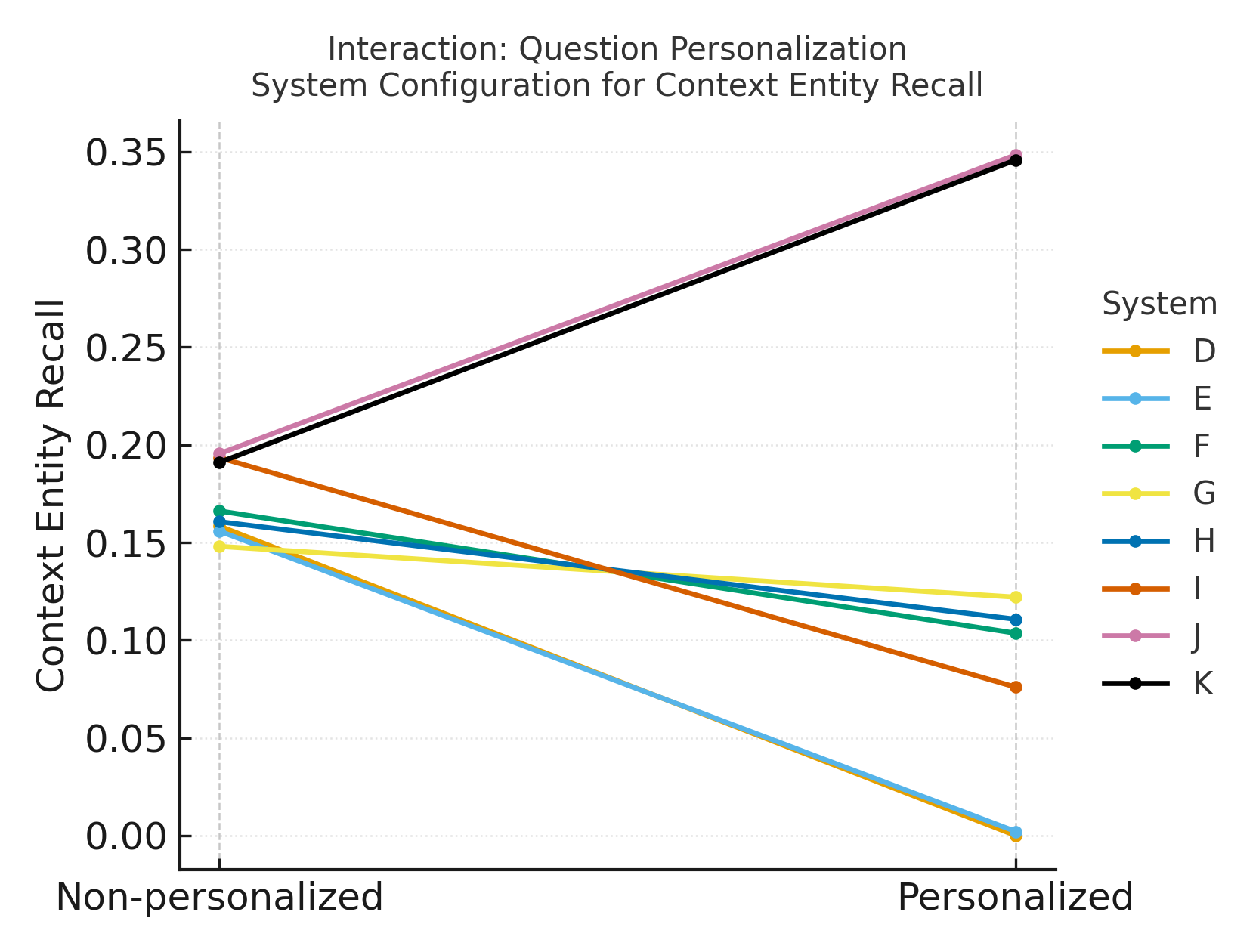} &
	\includegraphics[width=0.31\linewidth]{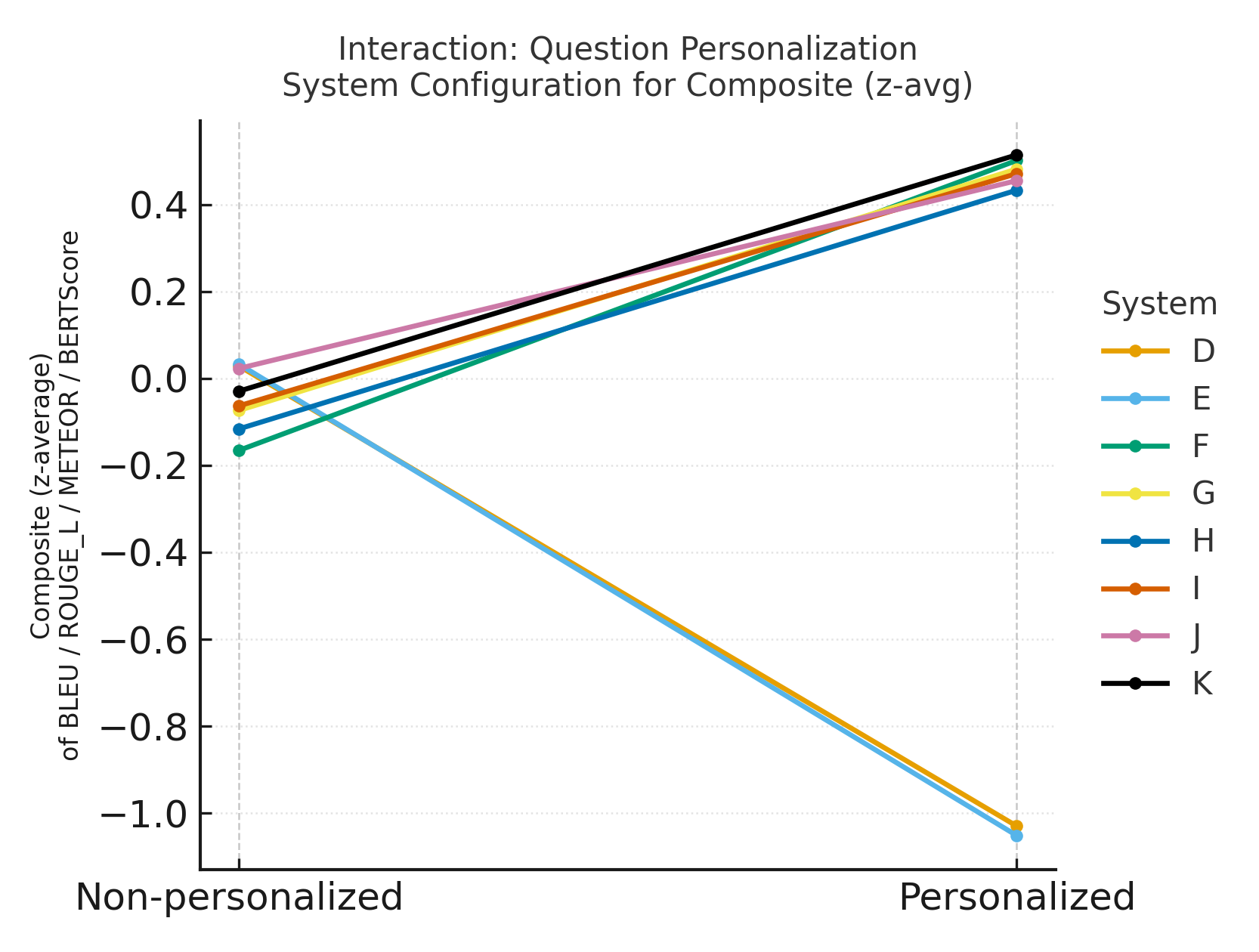} \\
	\end{tabular}
	\caption{Interaction plots by metric. Each panel shows system-level traces (A-K) across non-personalized and personalized questions with the legend placed at right within each panel. Panels include BLEU, ROUGE-L, METEOR, BERTScore, Faithfulness, Answer Relevancy, Answer Correctness, Context Entity Recall, and Composite (z-average of BLEU/ROUGE-L/METEOR/BERTScore).}
	\label{fig:interaction_grid_main}
\end{figure*}

This complex trade-off is visualized in the interaction plots shown in Figure~\ref{fig:interaction_grid_main}. These plots illustrate the performance of each system configuration across different metrics. The plots for Faithfulness and Answer Correctness clearly show an upward trend for the personalized, integrated systems (e.g., K, I, J). Conversely, the plot for BERTScore shows a distinct \emph{downward} or flat trend for these same systems, visually confirming the semantic loss penalty identified in the LMM analysis. This figure demonstrates that the strong gains in reasoning and grounding metrics were substantial enough to offset the semantic penalty, leading to an overall upward trend in the \texttt{Composite\_z} score.

\section{Discussion and Implications}

The AiVisor experiment provides critical evidence of the complex trade-offs in agentic personalization, moving beyond the field's assumption of uniform benefit. The primary LMM analysis revealed that personalization's effects are not uniformly positive; instead, it creates a trade-off between reasoning and semantic similarity. The analysis found significant negative interactions between role-framing and personalization factors on semantic metrics (e.g., BERTScore), alongside positive interactions on reasoning metrics (e.g., RAGAS Answer Correctness). This result, which may seem counter-intuitive, is directly explained by the experimental design. The test was systematically disadvantaged against personalization, using a question set ``skewed toward lexical precision'' and evaluating against a single, generic ground truth. Personalization *succeeds* by deviating from this generic reference to be more user-specific. Therefore, the observed semantic loss is a methodological artifact of standard metrics penalizing personalization for correctly doing its job. The fact that reasoning and grounding scores improved simultaneously confirms that this deviation was beneficial, not harmful. These results contribute methodologically by demonstrating that multi-locus personalization produces complex interactions. Theoretically, the findings support two conclusions: first, that standard evaluation benchmarks are insufficient for personalized Q\&A because they fail to account for beneficial deviation from a generic reference, and second, that personalization acts as a critical constraint on both evidence selection (retrieval) and expression generation. 

\subsection{Personalization Redundancy: Systems I, J, and K}

Comparison among systems I, J, and K isolates the impact of redundancy and role prompting. System I (prompt-level personalization with a role) exemplifies the central trade-off identified in the LMM analysis: it produces significant reasoning gains while incurring a significant semantic penalty. System J (retrieval-level personalization) improves evidence relevance but may drift in pragmatic scope without role guidance. System K integrates both and adds role prompting, achieving a synergistic effect specifically on reasoning, as shown by the positive three-way interaction for \texttt{Reasoning\_z} (coef: $0.044$). The redundancy observed in K appears to form a feedback alignment between retrieval bias and generation framing, leading to superior Context Precision and Faithfulness gains. This dual-conditioning mechanism, or redundant personalization, can be understood as a lightweight form of constraint harmonization similar in concept to bi-encoder feedback loops where personalization acts to align the output expression with the specific retrieved context. This is computationally necessary in high-stakes domains like advising, where failure to apply specific constraints (e.g., student major) leads to high-cost algorithmic errors \cite{Li2025}.

Redundant personalization introduces risks of over-constraining the model, where user-specific cues may propagate excessively through both retriever and generator. To mitigate this, a selective mechanism should modulate redundancy by detecting when retrieval already provides highly focused context. In those cases, prompt-level personalization can be reduced to avoid excessively narrow or privacy-sensitive phrasing. Conversely, when retrieval context is diffuse, prompt personalization should remain active to maintain relevance and user alignment.

\subsection{Practical implications and implementation guidance}

The practical implications are direct for system design. Role prompting should be enabled by default to maintain tone and domain scope consistency. VectorDB grounding is essential for factual scaffolding, as originally proposed in retrieval-augmented generation frameworks \cite{Lewis2020RAG}. Personalization in retrieval should be applied when user attributes directly affect relevance, which can improve Context Precision and Answer Relevancy. Prompt-level personalization, as the LMM analysis revealed, is a key factor in the reasoning-semantic trade-off: it should be used to improve reasoning quality, which showed a positive interaction, but designers must be aware that this will likely incur a semantic penalty (a BERTScore loss) when evaluated against generic references. When both mechanisms operate together, this redundancy should be selectively managed to avoid propagating sensitive user attributes. In operational deployments, the hierarchy observed in this study implies the following: use System I when retrieval cannot be modified, use System J when retrieval must be user-aware but roles are homogeneous, and use System K when both retriever and generator can share secure state information. System K thus represents the configuration with the best composite performance, achieving the most effective balance between reasoning gains and semantic penalties.

Beyond advising, the methodological pattern observed here generalizes to other retrieval-augmented tasks where personalization or context adaptation plays a role, including educational tutoring, clinical decision support, and customer interaction systems. By demonstrating the complex trade-offs between reasoning gains and semantic penalties under conservative conditions, AiVisor establishes a baseline and a methodological framework for future research in robust and transparent personalization for agentic AI.

\section{Limitations and Future Work}

The generalizability of these trade-offs is constrained by the intentionally conservative test environment built on twelve lexically-strict questions. We strategically minimized the evaluation scope to isolate and maximize the detection of the semantic penalty, but full validation requires replication on a larger, multi-turn, human-judged suite to confirm external validity. Although paired tests, effect sizes, sensitivity checks, and re-weighting help mitigate this limitation, a larger question suite will be required for full validation. The task mix is intentionally skewed toward generic, lexically strict prompts to create a conservative, worst-case setting; results should therefore be re-examined on multi-turn and domain-specific evaluations. The chosen four-metric panel captures core aspects of lexical and semantic quality but remains incomplete, as human judgments, calibration scores, and instruction-following measures warrant inclusion. System generality is also limited by reliance on a single base model family and a fixed prompt template set, motivating replication across alternative architectures and personalization schemas.

The current chunking configuration, which reduces retrieval precision and weakens personalization signals, is a key limitation. This factor, combined with the intentionally "lexically-skewed" question set, created the conservative test environment. The LMM analysis confirmed the outcome of this design: improvements were observed in RAGAS-based reasoning metrics, but these occurred alongside a significant penalty in semantic metrics (like BERTScore). This penalty is interpreted as a direct methodological artifact of evaluating user-specific deviations against a single, generic reference. Future work will extend the question set, refine retrieval granularity, and incorporate human assessments of appropriateness and usefulness. This is essential for addressing the ethical challenges of algorithmic bias and transparency \cite{Siddique2023} by confirming that the algorithmic reasoning gain translates into trustworthy and fair guidance from a human perspective.

\section{Conclusion}\label{conclu}

AiVisor demonstrates that agentic personalization produces complex, metric-dependent trade-offs, not simple uniform gains, even under conditions unfavorable to personalization. The Linear Mixed-Effects Model analysis, necessitated by the fractional-factorial design, isolated a key conflict: personalization factors (especially when interacting with a role prompt) significantly improved reasoning quality while simultaneously incurring a statistically significant penalty on semantic similarity metrics like BERTScore. This semantic loss can therefore be interpreted as a methodological artifact of the conservative, personalization-adverse evaluation, where lexically-oriented prompts and coarse chunking penalize deviations from a generic reference. Among tested configurations, System K-combining all factors-achieved the highest composite score, driven by strong gains in reasoning and grounding that outweighed the semantic metric penalty. This supports the hypothesis that the semantic loss was not a true loss of quality. These outcomes constitute a conservative validation of personalization, showing it yields measurable benefits in the correctness of an answer, even as it challenges the method used to measure semantic similarity.

Beyond empirical findings, AiVisor advances personalization research through methodological rigor and transparency. The evaluation framework integrates lexical and semantic similarity metrics with advanced statistical testing, moving beyond small-scale pilots and perception studies to provide system-level evidence of personalization effects. The analysis employed per-metric z-scoring and composite aggregation, while a Linear Mixed-Effects Model was used as the primary statistical test. This mixed-model approach was necessary to de-conflict the complex, nested factors of the experimental design, thereby providing the most robust framework for analyzing personalization's true effect. Future work will confirm the dominance of reasoning factors by calculating a Weighted Composite Score using the LMM coefficients as weights.

Evaluation under lexical stringency strengthens external validity by testing personalization in settings where it is least advantaged, conditions under which BLEU and ROUGE-L penalties are most likely. The multi-metric, normalized design addresses comparability across metrics and reduces the risk of selective reporting, while item-level accountability highlights which questions benefit from the reasoning gains and which incur lexical or semantic penalties. This granularity enables the development of selective mechanisms that determine when personalization should be applied or withheld, supporting responsible deployment.

Overall, AiVisor serves both as a prototype platform for academic advising and as an experimental testbed for studying personalization as an independent variable. The findings establish that personalization's effects are not uniformly positive, but rather create a complex trade-off between reasoning gains and semantic penalties (when measured against a lexical reference), providing an empirical foundation for cautious, data-driven personalization in production assistants. Future research should expand task coverage, incorporate human ratings of appropriateness and usefulness, extend to multi-turn workflows, and formalize selective mechanisms that balance personalization benefits with reliability in real-world applications.


\printbibliography

\end{document}